\documentclass[aps, prd, showkeys, singlecolumn, nofootinbib, floatfix, superscriptaddress]{revtex4-2}
\usepackage{amssymb}
\usepackage{amsmath}
\usepackage{graphicx}
\usepackage{dcolumn}
\usepackage{bm}
\usepackage[frak=esstix]{mathalpha}
\usepackage[colorlinks=true,allcolors=blue]{hyperref}
\usepackage{xcolor, soul, ulem, xpatch, caption, subcaption}
\usepackage{dutchcal}

\begin{document}

\title{Hydrodynamic theories for a system of weakly 
self-interacting classical ultra-relativistic scalar particles: causality and stability}

\author{Caio V. P. de Brito}
\email{caio\_brito@id.uff.br}
\affiliation{Instituto de F\'{\i}sica, Universidade Federal Fluminense \\ Av.~Gal.~Milton Tavares de Souza, S/N, 24210-346, Gragoatá, Niter\'{o}i, Rio de Janeiro, Brazil}
\affiliation{Institute for Theoretical Physics, Goethe University, Max-von-Laue-Str.~1, D-60438 Frankfurt am Main,  Germany}
\author{Gabriel S. Rocha}
\email{gabrielsr@id.uff.br}
\affiliation{Department of Physics and Astronomy, Vanderbilt University, 1221 Stevenson Center Lane
Nashville, TN 37212, USA}
\affiliation{Instituto de F\'{\i}sica, Universidade Federal Fluminense \\ Av.~Gal.~Milton Tavares de Souza, S/N, 24210-346, Gragoatá, Niter\'{o}i, Rio de Janeiro, Brazil}
\author{Gabriel S. Denicol}
\email{gsdenicol@id.uff.br}
\affiliation{Instituto de F\'{\i}sica, Universidade Federal Fluminense \\ Av.~Gal.~Milton Tavares de Souza, S/N, 24210-346, Gragoatá, Niter\'{o}i, Rio de Janeiro, Brazil}

\begin{abstract}
We investigate the causality and stability of three different relativistic dissipative fluid-dynamical formulations emerging from a system of classical, ultra-relativistic scalar particles self-interacting via a quartic potential. For this particular interaction, all transport coefficients of Navier-Stokes, Bemfica-Disconzi-Noronha-Kovtun and second-order transient theories can be computed in analytical form. We first show that Navier-Stokes theory is acausal and unstable regardless of the matching conditions. On the other hand, BDNK theory can be linearly causal and stable for a particular set of matching choices that does not contain the so-called exotic Eckart prescription. In particular, using the Liénard-Chipart criterion, we obtain a set of sufficient conditions that guarantee the stability of the theory. Last, second-order transient hydrodynamic theory in Landau matching is shown to be linearly causal and stable. 
\end{abstract}

\maketitle

\section{Introduction}

Formulations of relativistic dissipative fluid dynamics must satisfy two fundamental properties, namely causality and linear stability around equilibrium. The former forbids that signals propagate in the fluid with a velocity that exceeds the speed of light, while the latter dictates that perturbations around a global equilibrium state must decrease with time. These properties must be imprinted in the partial differential equations describing the evolution of the system and are largely entangled. In particular, in the linear regime, causality conditions with respect to a background fluid at rest coincide with the stability constrains with respect to a moving background fluid \cite{Pu:2009fj, Brito:2021iqr}. In fact, the interplay between causality and stability in fluid-dynamical theories has been addressed in Refs.~\cite{Gavassino:2021owo, Wang:2023csj}. 

Over the last years, several authors have pursued thorough understanding of causality in relativistic fluid dynamics \cite{Pu:2009fj, Bemfica:2017wps, Brito:2020nou, Bemfica:2020zjp, Brito:2021iqr, Gavassino:2021cli, Gavassino:2021kjm, Gavassino:2023odx, Gavassino:2023qwl, Sammet:2023bfo, Mullins:2023tjg}. In particular, in Ref.~\cite{Bemfica:2020zjp}, it was shown that if a given causal theory is stable under perturbations around a non-rotating equilibrium state in the local rest frame, it is also stable in any other Lorentz frame. In Refs.~\cite{Gavassino:2021cli, Gavassino:2021kjm}, causal fluid-dynamical theories were shown to be necessarily consistent with thermodynamic stability.  Moreover, in Refs.~\cite{Gavassino:2023odx, Gavassino:2023qwl}, universal equivalences were shown to appear in causal and thermodynamically stable hydrodynamic theories near equilibrium.

Historically, the search for causal and stable relativistic dissipative hydrodynamic theories starts with the fact that the relativistic counterpart of the Navier-Stokes theory \cite{Eckart:1940te, landau:59fluid} is acausal \cite{pichon:65etude} and unstable \cite{Hiscock:1983zz}. Hence, it is unsuitable for the modeling of relativistic fluids, even though its non-relativistic analogue is widely employed in the corresponding regime \cite{batchelor1967introduction}. This can be schematically explained by the fact that time- and space-like derivatives appear at unequal footing in these equations -- Navier-Stokes equations are parabolic \cite{pichon:65etude}. This property can be traced back to the fact that, in such theory, only space-like derivatives are taken into account in the constitutive relations connecting dissipative currents and gradients of thermodynamic fields, i.e., 4-velocity, temperature and chemical potential. In other words, the fluid response in terms of dissipative currents occurs instantaneously and only in the presence of \textit{space} inhomogeneities on the aforementioned fields in the fluid rest frame. 

In the last decade, the formulation pioneered by Israel and Stewart \cite{Israel:1976tn, Israel:1979wp} has been widely employed in the modeling of heavy-ion collisions \cite{Gale:2013da, Heinz:2013th, Florkowski:2017olj}. In this formalism, the dissipative currents are considered as independent variables which obey relaxation-type equations of motion independent of the local conservation laws. This renders the theory causal and stable in the linear regime as long as the relaxation times and further transport coefficients satisfy fundamental constraints \cite{Hiscock:1983zz, Olson:1990rzl, Muronga:2003ta, Denicol:2008ha, Pu:2009fj, Brito:2020nou, Sammet:2023bfo}. Causality constraints have been recently generalized to the full non-linear regime for uncharged, non-vortical fluids \cite{Bemfica:2019cop, Bemfica:2020xym}. Then, the constraints involve not only transport coefficients, but also the hydrodynamic fields (e.g.~bulk viscous pressure, shear-stress tensor) themselves. Moreover, in Ref.~\cite{Shokri:2023rpp}, it was shown that new propagation modes emerge in Israel-Stewart-like theories for perturbations around inhomogeneous equilibrium configurations.

More recently, the Bemfica-Disconzi-Noronha-Kovtun (BDNK) theory of hydrodynamics has been proposed \cite{Bemfica:2017wps,Bemfica:2020zjp,Kovtun:2019hdm}. In this formulation, analogously to Navier-Stokes theory, the dissipative currents also obey constitutive relations. However, the fundamental difference is that these equations contain not only space-like derivatives, but also time-like ones. Another noteworthy feature is the agnostic employment of matching conditions, a basic ingredient of all dissipative theories of hydrodynamics, which defines temperature, chemical potential and 4-velocity out of equilibrium. Then, the corresponding equations of motion can be causal in the non-linear regime and linearly stable, also leading to constrains for the transport coefficients \cite{Bemfica:2019knx}. Indeed, causality implies that e.g.~dissipative corrections to the energy density must be non-zero, as well as the energy diffusion current. This requires the employment of matching conditions that are more general than the traditional prescriptions by Landau \cite{landau:59fluid} and Eckart \cite{Eckart:1940te}.

This work is a continuation of Ref.~\cite{Rocha:2023hts}, where transport coefficients for Navier-Stokes, BDNK, and second-order transient hydrodynamic theories have been analytically computed for a system of classical ultra-relativistic scalar particles weakly self-interacting through the $\varphi^{4}$ potential. The particular form of the cross-section of the system renders it possible to compute the spectrum of the linearized collision operator in exact form \cite{Denicol:2022bsq}. This immensely simplifies the derivations of the transport coefficients because the inversion of this operator becomes trivial. In this follow-up paper, we assess the causality and stability of the aforementioned fluid-dynamical theories in the linear regime. In particular for BDNK theory, we determine the constraints on the class of matching conditions imposed by causality and stability. 

This work is organized as follows. In Sec.~\ref{sec:hydro-gen-match}, we introduce the formulation of relativistic dissipative hydrodynamics in generic matching conditions. We then summarize the different hydrodynamic theories derived in Ref.~\cite{Rocha:2023hts}. After that, in Sec.~\ref{sec:caus-stab-analysis}, we discuss the linear causality and stability for Navier-Stokes, BDNK and transient hydrodynamic theories, respectively, and Sec.~\ref{sec:conc} concludes the text. Last, in Appendix \ref{apn:hilb}, we study the causality and stability of the Hilbert theory. Throughout this work, we adopt the mostly minus convention for the Minkowski metric tensor, $g_{\mu\nu} = \mathrm{diag}(+---)$, and make use of natural units, $c = \hbar = k_B = 1$.

\section{Relativistic fluid dynamics for general matching conditions}
\label{sec:hydro-gen-match}

The fundamental equations of fluid dynamics are the continuity equations that describe the conservation of particle number, energy and momentum,
\begin{equation}
\label{eq:cons-laws}
\partial_\mu N^\mu = 0, \ \ \partial_\mu T^{\mu\nu} = 0,
\end{equation}
with $N^\mu$ being the particle 4-current and $T^{\mu\nu}$ the energy-momentum tensor. Considering generic matching conditions \cite{Rocha:2021lze}, these conserved currents can be expressed as
\begin{subequations}
\begin{align}
N^\mu &= (n_0 + \delta n) u^\mu + \nu^\mu, \\
T^{\mu\nu} &= (\varepsilon_0 +\delta \varepsilon) u^\mu u^\nu - \Delta^{\mu\nu}(P_0 + \Pi) + h^\mu u^\nu + h^\nu u^\mu + \pi^{\mu\nu} ,\label{eq:cons-energ}
\end{align}
\end{subequations}
where $\varepsilon_0$, $n_0$ are the equilibrium energy density and particle density, respectively, with $\delta \varepsilon$ and $\delta n$ denoting their corresponding dissipative corrections. Furthermore, $P_0$ is the thermodynamic pressure, $\Pi$ is the bulk viscous pressure, $u^{\mu }$ is the normalized 4-velocity, $u^{\mu}u_{\mu }=1$, $\nu^{\mu }$ is the particle diffusion current, $h^\mu$ is the energy diffusion current and $\pi^{\mu \nu }$ is the shear-stress tensor. The definition of the local equilibrium is determined by matching conditions, which will be discussed in the next section. They will enable us to identify the fluid 4-velocity, the thermal potential, $\alpha \equiv \mu/T$ (with $\mu$ being the chemical potential), and the temperature, $T$, out of equilibrium. We also introduced the projection operator onto the 3-space orthogonal to $u^{\mu }$, $\Delta ^{\mu \nu} \equiv g^{\mu \nu}-u^\mu u^\nu$. All 4-vector and rank-2 tensor dissipative currents are orthogonal to the fluid 4-velocity,
\begin{equation}
u_\mu \nu^\mu = 0, \ \ 
u_\mu h^\mu = 0, \ \ 
u_\mu \pi^{\mu\nu} = 0.
\end{equation}

It is convenient to decompose the conservation equations for energy and momentum into their parallel and orthogonal components with respect to the fluid 4-velocity. This is accomplished by contracting the second equality in Eq.~\eqref{eq:cons-laws} with $u_\nu$ and $\Delta_\nu^\alpha$, respectively, leading to the following equations of motion:
\begin{subequations}
 \label{eq:basic-hydro-EoM}
\begin{align}
 \label{eq:hydro-EoM-n0}
 Dn_{0}+D\delta n + (n_{0}+\delta n) \theta + \partial_{\mu} \nu^{\mu} &= 0, \\
\label{eq:hydro-EoM-eps}
 D\varepsilon_{0}+D\delta \varepsilon + (\varepsilon_{0}+\delta \varepsilon + P_{0} + \Pi) \theta - \pi^{\mu \nu} \sigma_{\mu \nu} + \partial_{\mu}h^{\mu} + u_{\mu} Dh^{\mu} &= 0, \\
\label{eq:hydro-EoM-umu}
(\varepsilon_{0} + \delta \varepsilon + P_{0} + \Pi)Du^{\mu} - \nabla^{\mu}(P_{0} + \Pi) + h^{\mu} \theta + h^{\alpha} \Delta^{\mu \nu} \partial_{\alpha}u_{\nu} +  \Delta^{\mu \nu} Dh_{\nu} + \Delta^{\mu \nu} \partial_{\alpha}\pi^{\alpha}_{ \ \nu} &= 0,
\end{align}
\end{subequations}
where $D \equiv u^\mu \partial_\mu$ is the comoving time derivative, $\nabla^\mu \equiv \Delta^{\mu\nu} \partial_\nu$ is the 4-gradient operator, $\theta \equiv \partial_\mu u^\mu$ is the expansion rate, and  $\sigma^{\mu \nu} \equiv \Delta^{\mu \nu \alpha \beta} \partial_{\alpha} u_{\beta} $ is the shear tensor, with $\Delta^{\mu\nu\alpha\beta} \equiv (\Delta^{\mu\alpha}\Delta^{\nu\beta} + \Delta^{\mu\beta}\Delta^{\nu\alpha})/2 - \Delta^{\mu\nu} \Delta^{\alpha\beta}/3$ being the double symmetric, traceless projection operator with respect to $u^\mu$. 

\subsection{Hydrodynamic theories}
\label{subsec:hydro-theories}

The local conservation laws, Eqs.~\eqref{eq:basic-hydro-EoM}, comprise 5 equations for, in principle, 19 variables ($n_{0}$, $\varepsilon_{0}$, $\delta n$, $\delta \varepsilon$, $\Pi$, $u^{\mu}$, $\nu^{\mu}$, $h^{\mu}$, and $\pi^{\mu \nu}$, since $P_{0} = P_{0}(n_{0},\varepsilon_{0})$ is not an independent variable). Thus, further dynamical/constitutive relations must be provided for closure. The specific procedure employed to obtain such relations defines a relativistic dissipative theory of hydrodynamics. 
Over the next sections, we shall consider three different formulations, namely Navier-Stokes theory, BDNK theory and transient second-order fluid dynamics, for a system of weakly self-interacting classical massless particles, derived in Ref.~\cite{Rocha:2023hts}. Before discussing these theories, we shall digress about the definition of the local equilibrium state.

Traditionally, the prescriptions constructed by Landau \cite{landau:59fluid} and Eckart \cite{Eckart:1940te} are employed to define the local equilibrium state, or, equivalently, the thermodynamic variables $\alpha$, $\beta \equiv 1/T$, and $u^{\mu}$. In both approaches, $\alpha$ and $\beta$ are defined so that the total particle and energy densities follow the equilibrium equation of state, which implies that 
\begin{equation}
\label{eq:land-eck-scal}
\delta n \equiv 0, \quad \delta \varepsilon \equiv 0.    
\end{equation}
On the other hand, the definition of $u^{\mu}$ is different in these prescriptions. For Landau, $u^{\mu}$ is defined as the time-like normalized eigenvalue of the energy-momentum tensor, $T^{\mu}_{\ \nu} u^{\nu} = \varepsilon_{0} u^{\mu}$, which sets the energy diffusion current to zero
\begin{equation}
\label{eq:landau-vec}
 h^{\mu} \equiv 0,   
\end{equation}
whereas for Eckart $u^{\mu}$ is the 4-velocity of the matter current, $N^{\mu} \equiv n_{0} u^{\mu}$, which implies that 
\begin{equation}
\label{eq:eckart-vec}
\nu^{\mu} \equiv 0.    
\end{equation}

Alternative matching prescriptions have been employed in Refs.~\cite{Rocha:2021lze,Rocha:2022ind, Rocha:2023hts,Rocha:2023ths} in the context of kinetic theory. In this framework, the particle 4-current and the energy-momentum tensor are integrals over momentum space of the single-particle distribution function, $f_{\bf p}$, whose dynamics is given by the Boltzmann equation,
\begin{equation}
\label{eq:currents_kin}
N^{\mu} = \int dP \, p^{\mu} f_{\textbf{p}}, \hspace{0.2cm}
T^{\mu\nu} = \int dP \, p^{\mu}p^{\nu} f_{\textbf{p}}.
\end{equation}
When describing fluids, it is often convenient to decompose the single-particle distribution function into an equilibrium part, $f_{0{\bf p}}$, and a dissipative correction, $\delta f_{\bf p}$,  
\begin{equation}
\begin{aligned}
&
f_{\bf{p}} = f_{0\bf{p}}+\delta f_{\bf{p}}.
\end{aligned}    
\end{equation}
Then, matching conditions are imposed to define the local equilibrium distribution. In Refs.~\cite{Rocha:2021lze, Rocha:2022ind, Rocha:2023hts, Rocha:2023ths} this is done by considering the wide (but not complete) set
of matching conditions,
\begin{equation}
\label{eq:kinetic_match}
\begin{aligned}
    \int dP \, E^{q}_{\textbf{p}} \delta f_{\textbf{p}} \equiv 0, \, \, \, 
    \int dP \, E_{\textbf{p}}^s \delta f_{\textbf{p}} \equiv 0, \, \, \,
    \int dP \, E_{\textbf{p}}^{z} p^{\langle \mu \rangle} \delta f_{\textbf{p}} \equiv 0,
\end{aligned}
\end{equation}
where $p^{\langle \mu \rangle} \equiv \Delta^{\mu\nu} p_\nu$, while $q$, $s$, and $z$ are free parameters. These conditions reduce to the Landau matching prescription [Eqs.~\eqref{eq:land-eck-scal} and \eqref{eq:landau-vec}] for $q=1$, $s=2$, and $z=1$ and to the Eckart prescription [Eqs.~\eqref{eq:land-eck-scal} and \eqref{eq:eckart-vec}] when $q=1$, $s=2$, and $z=0$. Other values of $q$, $s$, and $z$ lead to alternative matching conditions which do not possess an intuitive physical interpretation. In general, the transport coefficients of all hydrodynamic theories may depend on $q$, $s$, and $z$, and, as a consequence, the choice of matching can affect their causality and stability. 

Next, we proceed to discuss the complementary constitutive/dynamical relations for the dissipative currents that must be coupled with the conservation laws in order to close the system of partial differential equations given in Eqs.~\eqref{eq:basic-hydro-EoM}. As already mentioned, we consider a system composed of classical ultra-relativistic scalar particles described by the following Lagrangian density
\begin{equation}
\label{eq:lag-dens}
\mathcal{L} = \frac{1}{2} \partial_{\mu} \varphi \ \partial^{\mu} \varphi
-
\frac{\lambda \varphi^{4}}{4!}.
\end{equation}
The specific form of the total cross-section for this interaction, at leading order in the coupling constant $\lambda$, reads 
\begin{equation}
\label{eq:cross-sec-phi4}
\begin{aligned}
& \sigma(\mathcal{s}, \Theta) =  \frac{\lambda^2}{64
\pi^{2} \mathcal{s}} \equiv \frac{g}{2\pi \mathcal{s}}, 
\end{aligned}    
\end{equation}
where $\mathcal{s}$ and $\Theta$ are the total energy and scattering angle in the center-of-momentum frame, respectively, and we defined $g \equiv \lambda^{2}/(32 \pi)$. In this case, the spectrum of the \textit{linearized} collision operator of the Boltzmann equation can be computed in analytical form \cite{Denicol:2022bsq}. It is then possible to obtain \textit{exact} expressions for the transport coefficients, as first demonstrated in Ref.~\cite{Rocha:2023hts}.

\subsubsection{Navier-Stokes theory}

Navier-Stokes theory is constructed based on the concept that space inhomogeneities in the thermodynamic fields generate a response of the fluid in the form of dissipative currents. In the context of kinetic theory, this is implemented by the Chapman-Enskog procedure \cite{enskog1917kinetische, chapman1916vi, deGroot:80relativistic, Denicol:2021}, in which the single-particle distribution function is expanded in powers of space-like gradients of the hydrodynamic variables. The Navier-Stokes equations then originate from the first-order truncation of this expansion, thus being often referred to as a first-order theory. In Ref.~\cite{Rocha:2023hts}, the following constitutive relations were analytically derived:
\begin{equation}
\label{eq:NS-eqs}
\delta n= 0, \, \, \,
\nu^\mu = z\frac{3}{g\beta^2} \nabla^{\mu}\alpha, \, \, \,
\delta \varepsilon = 0, \, \, \,
h^\mu = (z-1)\frac{12}{g\beta^3} \nabla^{\mu}\alpha, \, \, \
\pi^{\mu \nu}= \frac{96}{g \beta^3} \sigma^{\mu \nu}.    
\end{equation}

The dissipative corrections to the particle and energy density are identically zero, since only massless particles were considered, see Ref.~\cite{Rocha:2023hts} for details. This implies that these hydrodynamic equations do not carry any dependence on $q$ and $s$. The particle and energy diffusion currents depend on gradients of the thermal potential, while the shear-stress tensor depends on gradients of the fluid 4-velocity. In particular, if $z=0$ ($z=1$), these definitions are in fact consistent with Landau (Eckart) matching conditions.

\subsubsection{Bemfica-Disconzi-Noronha-Kovtun theory}

BDNK theory has recently emerged as a formulation of relativistic dissipative hydrodynamics containing only first-order derivatives that can be constructed to be causal and stable \cite{Bemfica:2019knx}. The novel ingredient of this theory is the inclusion of time-like derivatives in the constitutive relations for the dissipative currents. This feature, together with the employment of alternative matching conditions, can render the theory causal and well-posed \cite{Bemfica:2020zjp}. Indeed, this fact inspired the development of the more general matching prescriptions presented in Eq.~\eqref{eq:kinetic_match} in the context of kinetic theory. In such framework, BDNK theory is obtained by a generalization of the Chapman-Enskog procedure, as originally proposed in Ref.~\cite{Rocha:2022ind}. In Ref.~\cite{Rocha:2023hts}, this method was applied to a system of classical massless particles, yielding the following constitutive relations:
\begin{equation}
\label{eq:const-rel-BDNK-chap-8}
\begin{aligned}
& \Pi 
=  \frac{\chi}{3} \left( \frac{D\beta }{\beta} -  \frac{\theta}{3} \right),
\ \
 \delta n 
= \xi \left( \frac{D\beta }{\beta} -  \frac{\theta}{3} \right),
\ \
 \delta \varepsilon 
= \chi  \left( \frac{D\beta }{\beta} -  \frac{\theta}{3} \right),\\
& \nu^{\mu} 
= \varkappa \left( \frac{\nabla^{\mu} \beta}{\beta} +   D u^{\mu}\right), \ \
 h^{\mu} 
= \lambda \left( \frac{\nabla^{\mu} \beta}{\beta} +   D u^{\mu}\right),
    \\
&
\pi^{\mu \nu} = 2 \eta \sigma^{\mu \nu}.
\end{aligned}
\end{equation}
The transport coefficients  are given in analytical form by
\begin{equation}
\label{eq:transp-coeffs-BDNK-cap8}
\begin{aligned}
&
\xi =  \frac{12}{g\beta ^2} (q-1) (s-1), \quad 
\chi = \frac{36}{g\beta^3}(q-2)(s-2), \quad
\varkappa = \frac{12}{g\beta ^2}z, \quad
\lambda 
= 
\frac{48}{g\beta ^3}(z-1), \quad
\eta
= \frac{48}{g\beta ^3}.
\end{aligned}    
\end{equation}

As previously discussed, the major contrast in comparison with Navier-Stokes equations is the presence of time-like derivatives in the constitutive relations for the dissipative currents, with the exception of the shear-stress tensor, which has the same form in both formulations. In addition, the dissipative corrections to particle and energy density are not always zero, but instead depend explicitly on the matching conditions. We note that, for the interaction considered in this paper, \textit{all} non-equilibrium fields depend on derivatives of the temperature but not of the chemical potential. Moreover, it can be readily seen that if we take $(q,s,z) = (1,2,1)$ and $(q,s,z) = (1,2,0)$, we recover Landau and Eckart matching conditions, respectively.

\subsubsection{Transient theory}

Second-order theories of fluid dynamics have been widely employed in models to emulate the evolution of the quark-gluon plasma phase of ultra-relativistic heavy-ion collisions \cite{Schenke_2010, Schenke_2011, Paquet_2016, Denicol_2018}. The derivation of these theories can be pursued phenomenologically, from the second law of thermodynamics \cite{Israel:1976tn}, and also via microscopic calculations, in particular, from kinetic theory \cite{Israel:1979wp}. In Ref.~\cite{Rocha:2023hts}, transient second-order equations of motion for the dissipative currents for a system with cross section given by Eq.~\eqref{eq:cross-sec-phi4} were analytically derived using the method of moments \cite{Denicol:2021} within the order of magnitude truncation scheme \cite{struchtrup2004stable, Fotakis:2022usk} and assuming Landau matching conditions, leading to
\begin{subequations}
\label{eq:IS-eoms}
\begin{align}
\tau_{\nu} D\nu^{\langle \lambda \rangle} 
+
\nu^{\lambda}  
&=
\varkappa_{n} \nabla^{\lambda}\alpha
-
\delta_{\nu \nu} \nu^{\lambda}\theta 
- (\lambda_{\nu \pi}
\nabla_{\mu} \alpha + \tau_{\nu \pi}
\nabla_{\mu} P_{0})\pi^{\lambda \mu} 
+
\ell_{\nu \pi}\Delta^{\lambda}_{ \ \alpha}
\nabla_{\mu}\pi^{\alpha \mu} 
-
\frac{7}{5}\tau_{\nu} \sigma_{\mu}^{\ \lambda}    \nu^{\mu} 
-
\tau_{\nu} \omega_{\mu}^{\ \lambda}  \nu^{\mu}
, \\
\tau_{\pi} D\pi^{\langle \lambda \mu \rangle}
+
\pi^{\lambda \mu} 
&=
2 \eta  \sigma^{\lambda \mu}
+
\varphi_{8}  \nu^{\langle \lambda} \nu^{\mu \rangle}
-
\delta_{\pi \pi} \pi^{\lambda \mu} \theta
- 
\tau_{\pi \nu}  \nabla^{\langle \lambda} P_{0} \ \nu^{\mu \rangle} 
+
\ell_{\pi \nu}  
 \nabla^{\langle \lambda}  \nu^{\mu \rangle} 
+
\lambda_{\pi \nu}
\nabla^{\langle \lambda} \alpha \ \nu^{\mu \rangle}
\notag \\
&-
2 \tau_{\pi}
 \omega_{\nu}^{\ \langle \lambda} \pi^{\mu \rangle \nu} 
- 
 \tau_{\pi\pi}
\sigma_{\nu}^{\ \langle \lambda}  \pi^{\mu \rangle \nu} 
,
\end{align}
\end{subequations}
where expressions for the transport coefficients were obtained in exact form as follows\footnote{In Ref.~\cite{Rocha:2023hts}, the diffusion coefficient was originally defined as $\kappa_n$. In order to avoid confusion with the covariant wave number, $\kappa$, which will be introduced in the next section, in this work we adopt the notation $\varkappa_{n}$ instead.}
\begin{eqnarray}
\label{eq:IS-coeffs}
&& \tau_\nu = \frac{60}{g n_{0} \beta^{2}}, \ \
\varkappa_{n} = \frac{3}{g \beta^{2}}, \ \ 
\lambda_{\nu \pi} = \frac{3 \tau_{\nu} \beta}{40}, \ \ 
\tau_{\nu \pi} = \frac{\tau_{\nu} \beta}{80 P_{0}}, \ \ 
\ell_{\nu \pi} = \frac{\tau_{\nu} \beta}{40}, \ \ 
\delta_{\nu \nu} = \tau_{\nu}, \notag \\
&& \tau_{\pi} = \frac{72}{g n_{0} \beta^{2}}, \ \ 
\eta = \frac{48}{g\beta^{3}}, \ \ 
\varphi_{8} = \frac{4}{n_{0} \beta}, \ \
\delta_{\pi \pi} = \frac{4}{3}\tau_{\pi}, \ \ 
\tau_{\pi\pi} = 2 \tau_{\pi}, \ \ 
\tau_{\pi \nu} = - \frac{4}{3} \frac{\tau_{\pi}}{n_{0}}, \\
&& \ell_{\pi \nu} = -\frac{4}{3} \frac{\tau_{\pi}}{\beta}, \ \ 
\lambda_{\pi \nu} = \frac{2}{3} \frac{\tau_{\pi}}{\beta}. \notag
\end{eqnarray}

The product of the coupling terms between the particle diffusion and the shear-stress tensor, $\ell_{\nu \pi}$ and $\ell_{\pi\nu}$, is negative-definite, in agreement with the second law of thermodynamics \cite{Brito:2020nou}. Moreover, the diffusion and shear viscosity coefficients are identical to the ones calculated in Navier-Stokes theory. Therefore, the transient equations reduce to the Navier-Stokes equations within Landau matching conditions when the relaxation times are taken to zero, i.e., $\tau_\nu \to 0$ and $\tau_\pi \to 0$.

The main goal of this paper is to analyze whether these different formulations of hydrodynamics are linearly causal and stable around global equilibrium. This can be fully determined now that exact expressions for the transport coefficients are known. In particular, when it comes to BDNK theory, we wish to quantify the effect that the choice of matching conditions can have on the linear stability and causality of the theory. This shall be carefully addressed in the following section.

\section{Linear causality and stability analysis}
\label{sec:caus-stab-analysis}

We assess the causality and linear stability of the above-mentioned theories by considering \textit{small} perturbations on all fluid-dynamical fields (denoted by $\Delta$) around a global equilibrium state, so that
\begin{eqnarray}
\varepsilon \rightarrow \varepsilon_0 + \Delta\varepsilon_0 + \Delta\delta\varepsilon, \hspace{.1cm}
n \rightarrow n_0 + \Delta n_0 + \Delta\delta n,
\hspace{.1cm} \Pi \rightarrow \Delta \Pi,
\hspace{.1cm} u^\mu \rightarrow u^\mu_0 + \Delta u^\mu, \hspace{.1cm}
\nu^\mu \rightarrow \Delta \nu^\mu, \hspace{.1cm}
h^\mu \rightarrow \Delta h^\mu, \hspace{.1cm}
\pi^{\mu\nu} \rightarrow \Delta \pi^{\mu\nu}.
\end{eqnarray}
In this case, the linearized fluid-dynamical equations, Eqs.~\eqref{eq:basic-hydro-EoM}, become
\begin{subequations}
\label{eq:linear-cons-laws}
\begin{align}
D_0 \Delta n_0 + D_0 \Delta\delta n + n_{0} \nabla^{0}_\mu \Delta u^\mu + \nabla^{0}_{\mu} \Delta\nu^{\mu} &= \mathcal{O}(\Delta^2) \approx 0, \\
D_0 \Delta\varepsilon_0 + D_0 \Delta\delta \varepsilon + (\varepsilon_{0}+ P_{0}) \nabla^{0}_\mu \Delta u^\mu + \nabla^{0}_{\mu} \Delta h^{\mu} &= \mathcal{O}(\Delta^2) \approx 0, \\
(\varepsilon_{0} + P_{0}) D_0 \Delta u^{\mu} - \frac{1}{3} \nabla^{\mu}_{0}(\Delta \varepsilon_0 + \Delta\delta\varepsilon) + D_0 \Delta h^{\mu} + \nabla^{0}_{\alpha} \Delta\pi^{\mu\alpha} &= \mathcal{O}(\Delta^2) \approx 0,
\end{align}
\end{subequations}
where $\mathcal{O}(\Delta^2)$ denote terms of second order (or higher) in perturbations. Furthermore, we have introduced the comoving time derivative with respect to the background velocity, $D_0 \equiv u^\mu_0 \partial_\mu$, and the projected 4-gradient, $\nabla_{0}^\mu \equiv \Delta^{\mu\nu}_{0} \partial_\nu \equiv (g^{\mu\nu} - u^\mu_{0} u^\nu_{0}) \partial_\nu$. In addition, we employed the equation of state for a massless gas, $\Delta\varepsilon_0 = 3 \Delta P_0$ and expressed the bulk viscous pressure in terms of the non-equilibrium deviation of the energy density as $\Pi = \delta\varepsilon/3$, see Ref.~\cite{Denicol:2012cn}.

It is practical to express the linearized fluid-dynamical equations in Fourier space. We adopt the following convention for the Fourier transform
\begin{equation}
\tilde{X}(k^{\mu }) = \int d^{4}x\hspace{0.1cm}\exp \left( -ix_{\mu }k^{\mu}\right) X(x^{\mu }), \ \ \ \ 
X(x^{\mu }) = \int \frac{d^{4}k}{(2\pi )^{4}}\hspace{0.1cm}\exp \left(ix_{\mu }k^{\mu }\right) \tilde{X}(k^{\mu }), \label{eq:def-fourier-trans}
\end{equation}
where $k^{\mu}=(\omega ,\mathbf{k})$, with $\omega$ being the frequency
and $\mathbf{k}$ the wave vector. As proposed in Ref.~\cite{Brito:2020nou}, we define the following covariant variables, 
\begin{equation}
\Omega \equiv u_{0}^{\mu }k_{\mu }, \ \ 
\kappa ^{\mu } \equiv \Delta _{0}^{\mu \nu }k_{\nu },
\end{equation}
which correspond to the frequency and wave vector in the local rest frame of the background system, respectively. We also introduce the covariant wave number as $\kappa=\sqrt{-\kappa_\mu \kappa^\mu}$.

The linearized fluid-dynamical equations in Fourier space are then given by
\begin{subequations}
\begin{align}
\Omega \Delta \tilde{n}_0 + \Omega \Delta\tilde{\delta n} + n_{0} \kappa_\mu \Delta \tilde{u}^\mu + \kappa_{\mu} \Delta\tilde{\nu}^{\mu} &= 0, \\
\Omega \Delta\tilde{\varepsilon}_0 + \Omega \Delta\tilde{\delta \varepsilon} + (\varepsilon_{0}+ P_{0}) \kappa_\mu \Delta \tilde{u}^\mu + \kappa_{\mu} \Delta \tilde{h}^{\mu} &= 0, \\
(\varepsilon_{0} + P_{0}) \Omega \Delta \tilde{u}^{\mu} - \frac{1}{3} \kappa^{\mu} (\Delta \tilde{\varepsilon}_0 + \Delta\tilde{\delta\varepsilon}) + \Omega \Delta \tilde{h}^{\mu} +  \kappa_{\alpha} \Delta\tilde{\pi}^{\mu\alpha} &= 0.
\end{align}
\end{subequations}
It is convenient to write these equations in terms of dimensionless variables,
\begin{subequations}
\label{eq:resc-lin-cons-laws-fourier}
\begin{align}
\Omega \Delta \hat{\tilde{n}}_0 + \Omega \Delta\hat{\tilde{\delta n}} + \kappa_\mu \Delta \tilde{u}^\mu + \kappa_{\mu} \Delta\hat{\tilde{\nu}}^{\mu} &= 0, \\
\Omega \Delta\hat{\tilde{\varepsilon}}_0 + \Omega \Delta\hat{\tilde{\delta \varepsilon}} + \frac{4}{3} \kappa_\mu \Delta \tilde{u}^\mu + \kappa_{\mu} \Delta \hat{\tilde{h}}^{\mu} &= 0, \\
\frac{4}{3} \Omega \Delta \tilde{u}^{\mu} - \frac{1}{3} \kappa^{\mu} (\Delta \hat{\tilde{\varepsilon}}_0 + \Delta \hat{\tilde{\delta\varepsilon}}) + \Omega \Delta \hat{\tilde{h}}^{\mu} +  \kappa_{\alpha} \Delta\hat{\tilde{\pi}}^{\mu\alpha} &= 0, \label{eq:vector-lin-fourier-cons-laws}
\end{align}
\end{subequations}
where we have defined the rescaled variables in Fourier space, 
\begin{eqnarray}
\Delta \hat{\tilde{n}}_0 \equiv \Delta \tilde{n}_0/n_0, \ \
\Delta \hat{\tilde{\delta n}} \equiv \Delta \tilde{\delta n}/n_0, \ \ 
\Delta \hat{\tilde{\varepsilon}}_0 \equiv \Delta \tilde{\varepsilon}_0/\varepsilon_0, \ \ 
\Delta \hat{\tilde{\delta \varepsilon}} \equiv \Delta \tilde{\delta \varepsilon}/\varepsilon_0, \notag \\ 
\Delta \hat{\tilde{\nu}}^\mu \equiv \Delta \tilde{\nu}^\mu/n_0, \ \ 
\Delta \hat{\tilde{h}}^\mu \equiv \Delta \tilde{h}^\mu/\varepsilon_0, \ \ 
\Delta\hat{\tilde{\pi}}^{\mu\nu} \equiv \Delta \tilde{\pi}^{\mu\nu}/\varepsilon_0.
\end{eqnarray}

We remark that the transverse (orthogonal to $\kappa^\mu$) and longitudinal (parallel to $\kappa^\mu$) components of Eqs.~\eqref{eq:resc-lin-cons-laws-fourier} decouple, and the resulting equations can be solved independently. In this context, we define a projection operator in Fourier space onto the 3-space orthogonal to $\kappa^{\mu}$,
\begin{equation}
\Delta^{\mu\nu}_\kappa\equiv g^{\mu\nu}+\frac{\kappa^\mu \kappa^\nu}{\kappa^2}.\label{proj_fourier}
\end{equation}
Then, an arbitrary 4-vector, $A^\mu$, can be decomposed into its transverse and longitudinal components with respect to $\kappa^\mu$ as
\begin{equation}
A^\mu = A_\|\kappa^\mu+A^\mu_\bot,\label{rank1_fourier}
\end{equation}
with the longitudinal component being defined as $A_\|=-\kappa_\mu A^\mu/\kappa$ while the transverse component is $A^\mu_\bot=\Delta^{\mu\nu}_{\kappa}A_\nu$. An analogous procedure can be performed to decompose an arbitrary traceless rank two tensor, $B^{\mu\nu}$. In this case, it is first essential to introduce the double, symmetric, and traceless projection operator in Fourier space as
\begin{equation}
\Delta^{\mu\nu\alpha\beta}_{\kappa}=\frac{1}{2}\left(\Delta^{\mu\alpha}_{\kappa}\Delta^{\nu\beta}_{\kappa}+\Delta^{\mu\beta}_{\kappa}\Delta^{\nu\alpha}_{\kappa}\right)-\frac{1}{3}\Delta^{\mu\nu}_\kappa\Delta^{\alpha\beta}_{\kappa}.\label{dproj_fourier}
\end{equation}
Then, a traceless second-rank tensor, $B^{\mu\nu}$, can be expressed as
\begin{equation}
B^{\mu\nu} = B_{\|}\frac{\kappa^{\mu}\kappa^{\nu}}{\kappa^{2}}+\frac{1}{3} B_{\|}\Delta_{\kappa}^{\mu\nu}+B_{\bot}^{\mu}\frac{\kappa^{\nu}}{\kappa}+B_{\bot}^{\nu}\frac{\kappa^{\mu}}{\kappa}+B_{\bot}^{\mu\nu},\label{rank2_fourier}
\end{equation}
with the corresponding projections being defined as $B_{\|}\equiv\kappa_{\mu}\kappa_{\nu}B^{\mu\nu}/\kappa^{2}$, $B_{\bot}^{\mu}\equiv-\kappa^{\lambda}\Delta_{\kappa}^{\mu\nu}B_{\lambda\nu}/\kappa$, and $B_{\bot}^{\mu\nu}\equiv\Delta_{\kappa}^{\mu\nu\alpha\beta}B_{\alpha\beta}$.

At this point, we can express the rescaled linearized conservation laws in Fourier space, Eqs.~\eqref{eq:resc-lin-cons-laws-fourier}, in terms of the longitudinal and transverse degrees of freedom. First, we note that the scalar equations are already expressed in terms of the longitudinal degrees of freedom. Then, contracting Eq.~\eqref{eq:vector-lin-fourier-cons-laws} with $- \hat{\kappa}_\mu$, we obtain the remaining longitudinal equation, leading to  
\begin{subequations}
\label{eq:linear-hydro-eqs-fourier-long}
\begin{align}
\Omega \Delta \hat{\tilde{n}}_0 + \Omega \Delta \hat{\delta \tilde{n}} - \kappa \Delta \tilde{u}_{\|} - \kappa \Delta\hat{\tilde{\nu}}_{\|} &= 0, \\
\Omega \Delta \hat{\tilde{\varepsilon}}_0 + \Omega \Delta\hat{\delta \tilde{\varepsilon}} - \frac{4}{3} \kappa \Delta \tilde{u}_{\|} - \kappa \Delta \hat{\tilde{h}}_{\|} &= 0, \\
\frac{4}{3} \Omega \Delta \tilde{u}_{\|} - \frac{1}{3} \kappa (\Delta \hat{\tilde{\varepsilon}}_0 + \Delta\hat{\delta \tilde{\varepsilon}}) + \Omega \Delta \hat{\tilde{h}}_{\|} -  \kappa \Delta\hat{\tilde{\pi}}_{\|} &= 0. \label{eq:vector-linear-hydro-eq}
\end{align}
\end{subequations}

In order to obtain the equations for the transverse degrees of freedom, we project Eq.~\eqref{eq:vector-lin-fourier-cons-laws} with $\Delta^{\mu\nu}_\kappa$, leading to 
\begin{equation}
\frac{4}{3} \Omega \Delta \tilde{u}^{\mu}_\bot + \Omega \Delta \hat{\tilde{h}}^{\mu}_\bot - \kappa \Delta\hat{\tilde{\pi}}^{\mu}_\bot = 0. \label{eq:linear-hydro-eqs-fourier-trans}
\end{equation}
These are the rescaled longitudinal and transverse projections of the linearized conservation laws in Fourier space. Naturally, it is still necessary to impose relations for the dissipative currents for closure. 

In particular, we remark that the fully transverse shear-stress tensor, $\pi^{\mu\nu}_\bot$, decouples from the equations of motion for the energy density and fluid velocity and therefore will not be investigated. Furthermore, given that stability of a causal fluid-dynamical theory in the local rest frame implies that such formulation is also stable in \textit{every} Lorentz frame \cite{Bemfica:2020zjp, Wang:2023csj}, in this work we solely analyze the case of perturbations on a fluid at rest, where $\Omega = \omega$ and $\kappa^2 = k^2$. 

\subsection{Navier-Stokes theory}

The linear causality and stability of the relativistic Navier-Stokes equations have been investigated by several authors in the past for both Eckart and Landau matching conditions \cite{pichon:65etude, Hiscock:1985zz, Hiscock:1987zz}. In these studies, it was demonstrated that the theory displays non-physical instabilities around global equilibrium that are directly related to its acausal nature. For the sake of completeness, in this subsection we repeat these analyses considering a wider set of matching conditions \textit{and} the transport coefficients derived for a system of weakly self-interacting classical massless particles. 

In Fourier space, the rescaled non-trivial Navier-Stokes constitutive relations, Eqs.~\eqref{eq:NS-eqs}, become
\begin{equation}
\Delta \hat{\tilde{\nu}}^\nu = i \bar{\varkappa}_{n} \kappa^\mu \Delta \tilde{\alpha}, \hspace{.2cm} \Delta \hat{\tilde{h}}^\mu = i \bar{\lambda}_{h} \kappa^\mu \Delta \tilde{\alpha}, \hspace{.2cm} \Delta \hat{\tilde{\pi}}^{\mu\nu} = i \bar{\eta} \left( \kappa^\mu \Delta \tilde{u}^\nu + \kappa^\nu \Delta \tilde{u}^\mu - \frac{2}{3} \Delta^{\mu\nu}_0 \kappa^\lambda \Delta \tilde{u}_\lambda \right),
\end{equation}
where we have defined the following transport coefficients
\begin{equation}
\label{eq:NS-resc-coeff}
\bar{\varkappa}_{n} = z\frac{3}{g n_0 \beta^2}, \hspace{.2cm} \bar{\lambda}_{h} = (z-1)\frac{12}{g\beta^3 \varepsilon_0}, \hspace{.2cm} \bar{\eta} = \frac{48}{g \beta^3 \varepsilon_0}.
\end{equation}
The transverse and longitudinal projections of these currents read
\begin{subequations}
\begin{align}
\Delta \hat{\tilde{\nu}}^\mu_\bot = 0, \hspace{.2cm}  \Delta \hat{\tilde{h}}^\mu_\bot = 0, \hspace{.2cm} \Delta\hat{\tilde{\pi}}^{\mu}_\bot = i \bar{\eta} \kappa \Delta u^{\mu}_\bot, \label{eq:NS-eqs-trans} \\
\Delta \hat{\tilde{\nu}}_\| = i \bar{\varkappa}_{n} \kappa \Delta \tilde{\alpha}, \hspace{.2cm} \Delta \hat{\tilde{h}}_\| = i \bar{\lambda}_{h} \kappa \Delta \tilde{\alpha}, \hspace{.2cm} \Delta \hat{\tilde{\pi}}_{\|} = \frac{4}{3} i \bar{\eta} \kappa \Delta \tilde{u}_{\|}. \label{eq:NS-eqs-long}
\end{align}
\end{subequations}

From Eqs.~\eqref{eq:linear-hydro-eqs-fourier-trans} and \eqref{eq:NS-eqs-trans}, we obtain the following dispersion relation associated with the transverse modes of Navier-Stokes theory,
\begin{equation}
\label{eq:disp-rel-NS-transv}
\Omega = \frac{3}{4} i \bar{\eta} \kappa^2.
\end{equation}
We note that this is a diffusion-like, and thus parabolic, dispersion relation. In the context of special relativity, parabolic differential equations are mathematically pathological since they admit acausal solutions \cite{Denicol:2021}. In fact, when calculated in a Lorentz-boosted frame, the mathematical properties of these equations change, and the resulting equation has an additional (in this case, also unstable) solution in comparison to the original equation in a rest frame \cite{Denicol:2021}, a clearly non-physical feature.

To obtain the dispersion relation for the longitudinal modes, we first express the perturbations of particle and energy density in terms of perturbations of thermal potential and inverse temperature, 
\begin{subequations}
\label{eq:e-and-n-as-alpha-beta}
\begin{align}
\Delta \varepsilon_0 &= \left.\frac{\partial \varepsilon_0}{\partial \alpha_0}\right\vert_{\beta_0} \Delta\alpha +  \left.\frac{\partial \varepsilon_0}{\partial \beta_0}\right\vert_{\alpha_0} \Delta\beta = \varepsilon_0 \Delta \alpha - \frac{4 \varepsilon_0}{\beta_0} \Delta \beta \Longrightarrow \Delta \hat{\tilde{\varepsilon}}_0 = \Delta \tilde{\alpha} - \frac{4}{\beta_0} \Delta \tilde{\beta}, \\
\Delta n_0 &= \left.\frac{\partial n_0}{\partial \alpha_0}\right\vert_{\beta_0} \Delta\alpha +  \left.\frac{\partial n_0}{\partial \beta_0}\right\vert_{\alpha_0} \Delta\beta = n_0 \Delta \alpha - \frac{3}{\beta_0} n_0 \Delta \beta \Longrightarrow \Delta \hat{\tilde{n}}_0 = \Delta \tilde{\alpha} - \frac{3}{\beta_0} \Delta \tilde{\beta},
\end{align}
\end{subequations}
where we have used that, for a dilute classical gas, 
\begin{eqnarray}
\left.\frac{\partial \varepsilon_0}{\partial \alpha_0}\right\vert_{\beta_0} = \varepsilon_0, \hspace{.2cm}  \left.\frac{\partial \varepsilon_0}{\partial \beta_0}\right\vert_{\alpha_0} = - 4 \frac{\varepsilon_0}{\beta_0}, \hspace{.2cm} \left.\frac{\partial n_0}{\partial \alpha_0}\right\vert_{\beta_0} = n_0, \hspace{.2cm} \left.\frac{\partial n_0}{\partial \beta_0}\right\vert_{\alpha_0} = - \frac{3}{\beta_0} n_0.
\end{eqnarray}
Therefore, from Eqs.~\eqref{eq:linear-hydro-eqs-fourier-long}, \eqref{eq:NS-eqs-long} and \eqref{eq:e-and-n-as-alpha-beta}, the rescaled equations for the longitudinal degrees of freedom are given by
\begin{equation}
\left( 
\begin{array}{ccc}
\Omega - i \bar{\varkappa}_{n} \kappa^2 & - \frac{3}{\beta_0} \Omega & -\kappa \\ 
\Omega - i \bar{\lambda}_{h} \kappa^2 & - \frac{4}{\beta_0} \Omega & - \frac{4}{3} \kappa \\
- \frac{1}{3}\kappa + i \Omega \kappa \bar{\lambda}_{h} &  \frac{4}{3 \beta_0} \kappa & \frac{4}{3} \Omega -  \frac{4}{3} i \bar{\eta} \kappa^2
\end{array}%
\right) \left( 
\begin{array}{c}
\Delta\tilde{\alpha} \\ 
\Delta\tilde{\beta} \\
\Delta \tilde{u}_{\|} \\
\end{array}%
\right)=0.
\end{equation}
This equation admits non-trivial solutions as long as the determinant of the matrix on the left-hand side is zero, leading to the dispersion relation associated with the longitudinal modes,
\begin{equation}
\Omega^3 + i \Omega ^2 \kappa ^2  \left[ (3 \Bar{\lambda}_{h} -4 \Bar{\varkappa}_{n} )-\Bar{\eta}  \right] + \Omega \kappa^2 \left[\Bar{\eta}  \kappa ^2 (3 \Bar{\lambda}_{h} -4 \Bar{\varkappa}_{n})- \frac{1}{3}\right] - \frac{1}{3} i \kappa ^4 (3 \Bar{\lambda}_{h} -4 \Bar{\varkappa}_{n} ) = 0.
\end{equation}

We then analyze the solutions for a system with zero background velocity. In the large wave number limit, the modes behave as
\begin{equation}
\omega = i \bar\eta k^2 + \mathcal{O}(k), \ \
\omega = i (4 \bar\varkappa_{n} - 3 \bar\lambda_{h}) k^2 + \mathcal{O}(k), \ \
\omega = \frac{i}{3 \bar\eta} + \mathcal{O}\left( \frac{1}{k} \right).
\end{equation}
We note that, for perturbations on a static fluid, the modes are stable for all matching conditions since, $4 \bar\varkappa_{n} - 3 \bar\lambda_{h} >0$, see Eq.~\eqref{eq:NS-resc-coeff}. However, similarly to what was observed for the transverse modes, two of the longitudinal modes display a diffusion-like behavior for large values of $k$, hence leading to an additional unstable and acausal mode when considering perturbations on a moving fluid \cite{Denicol:2021}. Overall, we conclude that Navier-Stokes theory for weakly self-interacting classical massless particles is ill-defined and unsuitable to describe such systems regardless of the matching condition employed.

For the sake of completeness, in Appendix \ref{apn:hilb}, we show that Hilbert theory, a hydrodynamic theory whose microscopic derivation inspired the Chapman-Enskog expansion, does not display such instability and acausality problems. However, we remark that this formulation is usually not employed in practical applications since it requires a new conservation law at each order in the gradient expansion.

\subsection{BDNK theory}

The causality and linear stability of BDNK theory have been studied for several different scenarios \cite{Bemfica:2017wps, Bemfica:2019knx, Roy:2023apk, Rocha:2023hts,Armas:2022wvb}. Nevertheless, we emphasize that the theory derived for a system of weakly self-interacting classical massless particles, outlined in the previous section, has novel features -- in particular, the constitutive relations do not contain any time-like or space-like derivatives of the thermal potential. In this subsection, we shall investigate, for the first time, the stability and causality of this formulation with transport coefficients derived from the interaction described by Eq.~\eqref{eq:lag-dens}.

In Fourier space, the rescaled linearized BDNK equations, Eqs.~\eqref{eq:const-rel-BDNK-chap-8}, become
\begin{equation}
\begin{aligned}
& \Delta \hat{\tilde{\Pi}} 
=  i \frac{\bar{\chi}}{3} \left[ \Omega \left( \delta \hat{\tilde{n}} - \delta \hat{\tilde{\varepsilon}} \right) + \frac{\kappa \Delta \tilde{u}_\|}{3} \right],
\\
& \delta \Delta \hat{\tilde{n}} = i \bar{\xi} \left[ \Omega \left( \delta \hat{\tilde{n}} - \delta \hat{\tilde{\varepsilon}} \right) + \frac{\kappa \Delta \tilde{u}_\|}{3} \right],
\ \
\delta \Delta \hat{\tilde{\varepsilon}} = i \bar{\chi} \left[ \Omega \left( \delta \hat{\tilde{n}} - \delta \hat{\tilde{\varepsilon}} \right) + \frac{\kappa \Delta \tilde{u}_\|}{3} \right],\\
& \Delta \hat{\tilde{\nu}}^{\mu} 
= i \bar{\varkappa} \left[ \kappa^\mu \left( \delta \hat{\tilde{n}} - \delta \hat{\tilde{\varepsilon}} \right) + \Omega \Delta \tilde{u}^\mu \right], \ \
\Delta \hat{\tilde{h}}^{\mu} 
= i \bar{\lambda} \left[ \kappa^\mu \left( \delta \hat{\tilde{n}} - \delta \hat{\tilde{\varepsilon}} \right) + \Omega \Delta \tilde{u}^\mu \right],
    \\
&
\Delta \hat{\tilde{\pi}}^{\mu \nu} = i \bar{\eta} \left( \kappa^\mu \Delta \tilde{u}^\nu + \kappa^\nu \Delta \tilde{u}^\mu - \frac{2}{3} \Delta^{\mu\nu}_0 \kappa^\lambda \Delta \tilde{u}_\lambda \right),
\end{aligned}
\end{equation}
where we have defined the rescaled transport coefficients following the convention adopted in the last section,
\begin{equation}
\label{eq:BDNK-resc-coeff}
\begin{aligned}
&
\bar{\xi}=  \frac{12}{g\beta ^2 n_0} (q-1) (s-1), \quad 
\bar{\chi} = \frac{12}{g\beta^2 n_0}(q-2)(s-2),
\\
&
\bar{\varkappa} = \frac{12}{g\beta ^2 n_0}z, \quad 
\bar{\lambda} = \frac{16}{g\beta ^2 n_0}(z-1),
\\
&
\bar{\eta}
= \frac{16}{g\beta ^2 n_0},
\end{aligned}    
\end{equation}
and made use of the identity $\Delta \beta/\beta_0 = \Delta n/n_0 - \Delta \varepsilon/\varepsilon_0$, which stems from Eqs.~\eqref{eq:e-and-n-as-alpha-beta}.

Once again, we shall begin with the analysis of the transverse modes. First, we calculate the transverse projection of the energy diffusion and shear-stress tensor, 
\begin{equation}
\Delta \hat{\tilde{h}}^\mu_\bot = i \bar{\lambda} \Omega \Delta u^{\mu}_\bot, \hspace{.2cm} \Delta\hat{\tilde{\pi}}^{\mu}_\bot = i \bar{\eta} \kappa \Delta u^{\mu}_\bot.
\end{equation}
Substituting these relations in Eq.~\eqref{eq:linear-hydro-eqs-fourier-trans}, we obtain the transverse dispersion relation
\begin{equation}
\label{eq:BDNK-long-trans}
i \bar{\lambda} \Omega^2 + \frac{4}{3} \Omega - i \bar{\eta} \kappa^2 = 0. 
\end{equation}
Note that BDNK equations do not lead to a parabolic dispersion relation, which is the source of the non-physical features displayed by Navier-Stokes theory, as discussed in the previous section [cf.~Eq.~\eqref{eq:disp-rel-NS-transv}], as a consequence of the inclusion of time-like derivatives in the constitutive relation for the energy diffusion, $h^{\mu}$. 

Considering perturbations on a fluid at rest, the solutions of the transverse dispersion relation read
\begin{equation}
\omega^\pm = \frac{i}{3 \bar{\lambda}} \left(2 \pm \sqrt{4 - 9 \bar{\eta} \bar{\lambda} k^2} \right).
\end{equation}
If the term inside the square root is non-negative, both transverse modes are purely imaginary and positive-definite as long as $\lambda > 0$, which further implies that $z>1$. This is a necessary condition for stability and it was first derived in Ref.~\cite{Rocha:2023hts} considering homogeneous perturbations, i.e., $k=0$, with a similar analysis being pursued in Ref.~\cite{Roy:2023apk}. Therefore, the stability of BDNK theory depends on the choice of matching conditions. Since $\eta > 0$, this term is either positive and smaller than 2, or negative, in that case yielding modes that are oscillating as well as damping, if $k > 1/(3\sqrt{\bar{\eta} \bar{\lambda}})$. In both cases, the modes are stable for any value of $k$.

\begin{figure}[htb]
    \centering
    \begin{subfigure}[b]{\textwidth}
        \centering
        \includegraphics[scale=0.475]{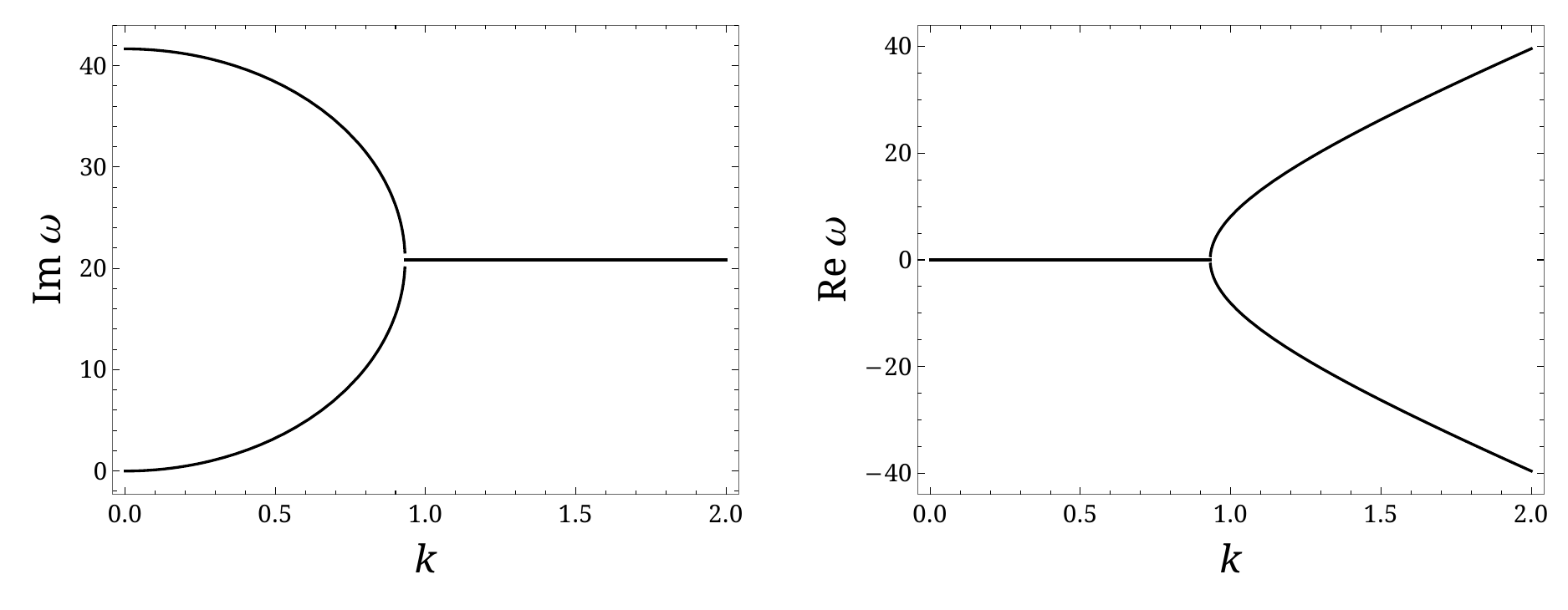}
        \hfill
        \caption{$z=1.002$}
        \label{fig:BDNK-trans-stable}
    \end{subfigure}
    \vskip \baselineskip
    \begin{subfigure}[b]{\textwidth}
        \centering
        \includegraphics[scale=0.475]{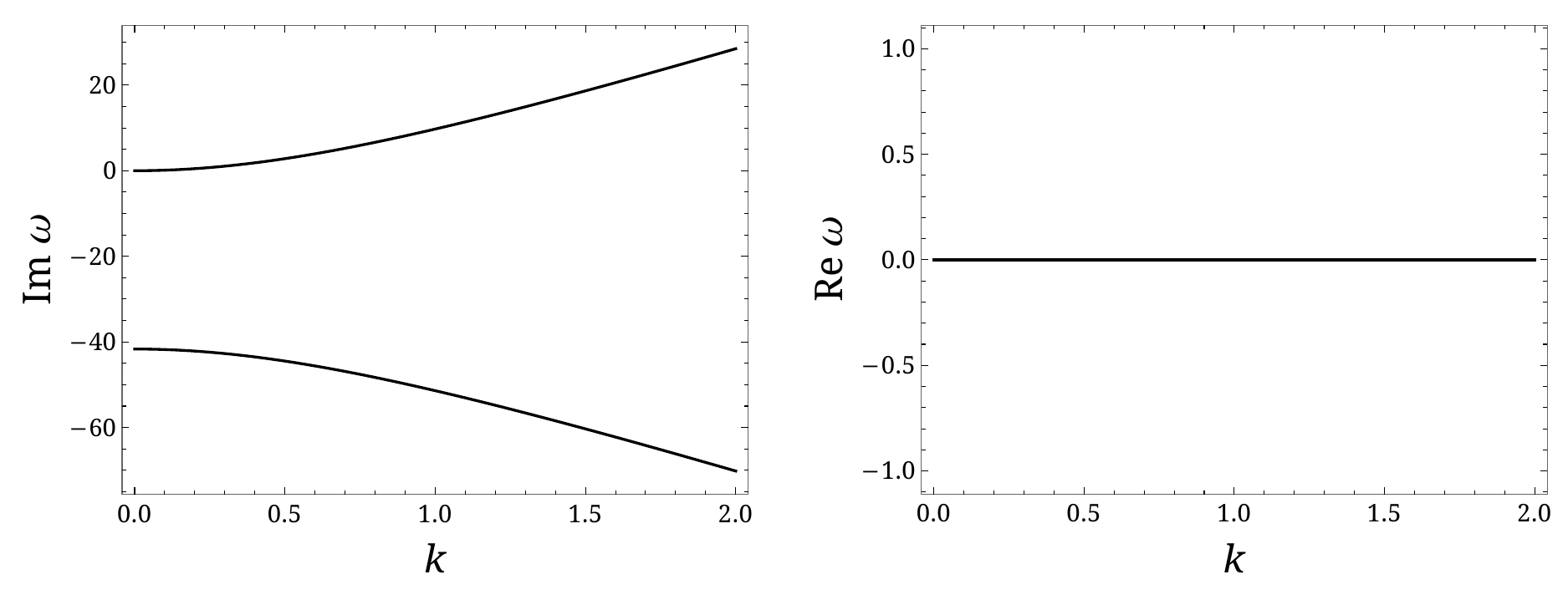}
        \hfill
        \caption{$z=0.998$}
        \label{fig:BDNK-trans-unstable}
    \end{subfigure}
    \caption{Imaginary and real part of the transverse modes of BDNK theory for two values of $z$.}
\end{figure}

In Fig.~\ref{fig:BDNK-trans-stable}, we portray the transverse modes of BDNK theory assuming $z = 1.002$, satisfying the stability condition obtained above. Indeed, it can be readily seen that the imaginary part of both modes is positive for any value of wave number, and therefore such modes are stable. In Fig.~\ref{fig:BDNK-trans-unstable}, on the other hand, we display the transverse modes assuming $z=0.998$, which violates the stability condition. As expected, one of the modes has a negative imaginary part, thus being unstable.

The next step is to analyze the longitudinal modes of the theory. The longitudinal projections of the dissipative currents are given by
\begin{equation}
\Delta \hat{\tilde{\nu}}_{\|} = i \bar{\varkappa} \left[ \kappa \left( \Delta \hat{\tilde{n}}_0 - \Delta \hat{\tilde{\varepsilon}}_0 \right) + \Omega \Delta \tilde{u}_{\|} \right], \ \
\Delta \hat{\tilde{h}}_{\|} = i \bar{\lambda} \left[ \kappa \left( \Delta \hat{\tilde{n}}_0 - \Delta \hat{\tilde{\varepsilon}}_0 \right) + \Omega \Delta \tilde{u}_{\|} \right], \ \
\Delta \hat{\tilde{\pi}}_{\|} = \frac{4}{3} i \bar{\eta} \kappa \Delta \tilde{u}_{\|}.
\end{equation}
Substituting these relations into Eqs.~\eqref{eq:linear-hydro-eqs-fourier-long} and expressing the resulting equations in matrix form, we obtain 
\begin{equation}
\left( 
\begin{array}{ccc}
\Omega + i \bar{\xi} \Omega^2 - i \bar{\varkappa} \kappa^2 & - i \bar{\xi} \Omega^2 +  i \bar{\varkappa} \kappa^2 & -\kappa + \frac{i}{3} \bar{\xi} \Omega \kappa - i \bar{\varkappa} \Omega \kappa  \\ 
i \bar{\chi} \Omega^2 - i \bar{\lambda} \kappa^2 & \Omega - i \bar{\chi} \Omega^2  + i \bar{\lambda} \kappa^2 & - \frac{4}{3} \kappa + \frac{i}{3} \bar{\chi} \Omega \kappa - i \bar{\lambda} \Omega \kappa \\
- \frac{i}{3} \bar{\chi} \Omega \kappa + i \bar{\lambda} \Omega \kappa & - \frac{\kappa}{3} + \frac{1}{3} i \bar{\chi} \Omega \kappa - i \bar{\lambda} \Omega \kappa & \frac{4}{3} \Omega -  \frac{4}{3} i \bar{\eta} \kappa^2 - \frac{i}{9} \bar{\chi}  \kappa^2 + i  \bar{\lambda} \Omega^2
\end{array}%
\right) \left( 
\begin{array}{c}
\Delta \hat{\tilde{n}} \\ 
\Delta \hat{\tilde{\varepsilon}} \\
\Delta \tilde{u}_{\|} \\
\end{array}%
\right)=0.
\end{equation}
Once again, this equation admits non-trivial solutions only if the determinant of the matrix on the left-hand side is zero, leading to a dispersion relation whose solutions are the longitudinal modes of the theory
\begin{eqnarray}
\label{eq:BDNK-long-disp}
\Omega ^5 \bar\lambda (\bar\chi - \bar\xi)+\frac{i}{9} \Omega ^4 (9 \bar\lambda +12 \bar\xi -12 \bar\chi ) 
+
\frac{1}{9} \Omega ^3 \left[6 \kappa ^2 (2 \bar\eta + \bar\lambda ) (\bar\xi - \bar\chi) + 12\right]
-
\frac{2}{9} i \kappa ^2 \Omega ^2 (6 \bar\eta -3 \bar\lambda +2 \bar\xi +6 \bar\varkappa -2 \bar\chi )
\notag 
\\
+
\frac{1}{9} \Omega  \left[\kappa ^4 \bar\lambda  (12 \bar\eta - \bar\xi + \bar\chi )-12 \bar\eta \bar\varkappa \kappa^4 - 4 \kappa ^2\right] 
+
\frac{i}{9} \left(4 \bar\varkappa - 3 \bar\lambda \right) \kappa ^4  = 0.   
\end{eqnarray}
At small wave number, these modes become
\begin{subequations}
\begin{align}
\omega_{\mathrm{nh}} &= \frac{4i}{3 \bar{\lambda}} + \mathcal{O}(k^2), \\
\omega_{\mathrm{nh}} &= \frac{i}{\bar{\xi} - \bar{\chi}} + \mathcal{O}(k^2), \\
\omega_{\mathrm{h}}  &= \pm \frac{1}{\sqrt{3}} \, k + \mathcal{O}(k^2), \\
\omega_{\mathrm{h}}  &= \frac{i}{4} (4 \bar{\varkappa} - 3 \bar{\lambda}) \, k^2 + \mathcal{O}(k^3),
\end{align}
\end{subequations}
where `n' stands for hydrodynamic and `nh' stands for non-hydrodynamic, i.e., modes that either do or do not vanish at $k \to 0$, respectively. The imaginary part of these modes are positive as long as the transport coefficients satisfy
\begin{equation}
\label{eq:BDNK-stab-long-small-k}
\bar\lambda > 0, \ \ \bar\xi > \bar\chi, 
\end{equation}
which further constrain the matching parameters as follows
\begin{equation}
\label{eq:stab-cond-bdnk-1}
z > 1, \ \ q+s > 3.
\end{equation}
We remark that these conditions have been first obtained in Ref.~\cite{Rocha:2023hts} for the case where the fluid-dynamical perturbations are homogeneous.

Moreover, in the large wave number limit, the longitudinal modes become
\begin{subequations}
\begin{align}
\omega &= \pm k \sqrt{\frac{1}{3} + \frac{2}{3\lambda}  \left(
\bar{\eta} \pm \sqrt{\frac{\bar\eta}{\bar\xi - \bar\chi} \left[\bar\eta  \bar\xi + 3 \bar\lambda ^2 + \bar\lambda  \bar\xi - 3 \bar\lambda  \bar\varkappa - \bar\chi  (\bar\eta + \bar\lambda ) \right]}\right)} + \mathcal{O}\left( \frac{1}{k} \right), \label{eq:bdnk-large-k-mode1} \\
\omega &= \frac{ (3 \bar{\lambda} -4 \bar{\varkappa}) i}{12 \bar{\eta}  (\bar\lambda - \bar{\varkappa}) + \bar{\lambda}( \bar{\chi} - \bar{\xi}) } + \mathcal{O}\left( \frac{1}{k} \right). \label{eq:bdnk-large-k-mode2}
\end{align}
\end{subequations}
First, stability implies that both outer and inner square roots in Eq.~\eqref{eq:bdnk-large-k-mode1} must be real, otherwise leading to an unstable mode. In particular, we have previously imposed that $\bar{\xi} - \bar\chi > 0$, Eq.~\eqref{eq:BDNK-stab-long-small-k}, hence rendering the numerator in the inner square root positive. In order for stability to be satisfied, the numerator must be identically positive. Since $\eta$ is positive-definite, we have
\begin{equation}
\bar\eta  \bar\xi + 3 \bar\lambda ^2 + \bar\lambda  \bar\xi - 3 \bar\lambda  \bar\varkappa - \bar\chi  (\bar\eta + \bar\lambda ) \geq 0,
\end{equation}
which leads to 
\begin{equation}
q+s\geq \frac{-z^2+8 z-4}{z}.
\label{eq:stab-cond-bdnk-q+s-1}
\end{equation}
On top of that, we must also ensure the term inside the outer square root is positive, and thus real, leading to
\begin{equation}
12 \bar\eta  (\bar\varkappa - \bar\lambda) + \bar\lambda (\bar\xi - \bar\chi) \geq 0.
\end{equation}
In terms of the matching parameters, this reduces to
\begin{equation}
q+s \geq \frac{7z -19}{z-1}.
\label{eq:stab-cond-bdnk-q+s-2}
\end{equation}
In particular, since $3 \bar\lambda - 4 \bar\varkappa < 0$, the stability of the mode given in Eq.~\eqref{eq:bdnk-large-k-mode2} implies that the denominator is identically negative, hence excluding the equality in Eq.~\eqref{eq:stab-cond-bdnk-q+s-2} and providing a stronger constraint. Furthermore, we remark that the stability condition given by Eq.~\eqref{eq:stab-cond-bdnk-q+s-2} surmounts the one given in Eq.~\eqref{eq:stab-cond-bdnk-q+s-1} if $z>4$. 

Furthermore, causality implies that the asymptotic group velocity must be smaller than the speed of light, 
\begin{equation}
\lim_{k\rightarrow \infty }\left\vert\frac{\partial\mathrm{Re} \, \, \omega}{\partial k}\right\vert \leq1.
\label{eq:def-causality}
\end{equation}
Then, causality is guaranteed as long as the term inside the outer square root in Eq.~\eqref{eq:bdnk-large-k-mode1} is smaller than $1$, leading to the following condition,
\begin{equation}
\bar{\lambda} - \bar{\eta} \geq \sqrt{\frac{\bar\eta}{\bar\xi - \bar\chi} \left[\bar\eta  \bar\xi + 3 \bar\lambda ^2 + \bar\lambda  \bar\xi - 3 \bar\lambda  \bar\varkappa - \bar\chi  (\bar\eta + \bar\lambda ) \right]}.
\end{equation}
Stability then dictates that
\begin{equation}
\lambda \geq \eta,
\end{equation}
which leads to $z \geq 2$. Then, we finally obtain the causality condition for the BDNK theory,
\begin{equation}
\label{eq:bdnk-causal-transp}
3 \bar\eta  (-\bar\lambda - \bar\xi + \bar\varkappa + \bar\chi) + \bar\lambda (\bar\xi - \bar\chi) \geq 0.
\end{equation}
This condition can be expressed in terms of the matching parameters as follows
\begin{equation}
\label{eq:bdnk-causal-params}
(q+s-4) (z-4) \geq 0.
\end{equation}

\subsection*{Liénard-Chipart criterion}

So far, we solved the dispersion relations and obtained the transverse and longitudinal modes of BDNK theory. In general, since these solutions are rather cumbersome, we resorted to analyzing the asymptotic behavior of the modes, i.e., in the small ($k \to 0$) and large ($k \to \infty$) wave number limits. We then required that all modes have positive imaginary parts and their real parts are not greater than 1 when $k \to \infty$. This provided us with necessary constraints for the transport coefficients, and, consequently, for the matching parameters, to ensure that stability and causality are simultaneously satisfied. 

We remark, however, that these conditions might still not be sufficient to guarantee that stability is always satisfied, since we lack the knowledge on how the modes behave at intermediate values of wave number. In particular, we can obtain a set of \textit{necessary} stability conditions using the Routh-Hurwitz criterion \cite{routh1877, hurwitz1895, korn2000}. The criterion gives the number of roots with positive real parts of a polynomial of the type
\begin{equation}
\label{eq:RH-polyn}
f(\Upsilon) = \sum_{j=0}^{N} a_{N-j} \Upsilon^{j}.     
\end{equation}
Defining $\Upsilon \equiv i \omega$, so that $\Re (\Upsilon) = - \Im (\omega)$ and expressing the dispersion relations, Eqs.~\eqref{eq:BDNK-long-trans} and \eqref{eq:BDNK-long-disp}, as polynomials of $\Upsilon$, the Routh-Hurwitz criterion gives the number of modes with negative imaginary parts. In this case, stability is guaranteed if, and only if, this number is exactly zero, i.e., if all modes have positive imaginary real parts. In this case, the polynomial is said to be Hurwitz stable. 

The Hurwitz stability of a real polynomial of the type given in Eq.~\eqref{eq:RH-polyn} is guaranteed if, and only if, the sequence consisting of $a_{0}$ and \textit{all} the principal minor determinants of the Hurwitz matrix associated with the polynomial above are positive. The Hurwitz matrix is a square matrix of dimension $N$ whose elements are obtained from the coefficients $a_{j}$ in Eq.~\eqref{eq:RH-polyn},
\begin{equation}
\begin{aligned}
&
H = \left( 
\begin{array}{ccccc}
a_{1}  &  a_{0} & 0 & \cdots & 0 \\
a_{3}  &  a_{2} & a_{1} & \cdots & 0 \\
a_{5}  &  a_{4} & a_{3} & \cdots & 0 \\
\vdots & \vdots  &  \vdots &  \vdots & \vdots
\\
a_{2\lfloor N/2 \rfloor - 1} & a_{2\lfloor N/2 \rfloor - 2 } & a_{2\lfloor N/2 \rfloor - 3} & \cdots & a_{\lfloor N/2 \rfloor} \\
0 & 0 & a_{2\lfloor N/2 \rfloor - 1} & \cdots & a_{\lfloor N/2 \rfloor + 2}
\\
\vdots & \vdots  &  \vdots &  \vdots & \vdots
\\
0 & 0 & 0 & \cdots & a_{2\lfloor N/2 \rfloor - 1}
\end{array} 
\right),
\end{aligned}   
\end{equation}
where $\lfloor N/2 \rfloor$ denotes the smallest integer larger than, or equal to $N/2$. The principal minor determinants of $H$ read
\begin{equation}
\begin{aligned}
&
T_{1} = a_{1}, 
\quad
T_{2} = 
\det\left( 
\begin{array}{cc}
a_{1} & a_{0} \\
a_{3}  & a_{2}
\end{array}
\right),
\quad
T_{3} = \det\left( 
\begin{array}{ccc}
a_{1} & a_{0} & 0 \\
a_{3}  & a_{2} & a_{1} \\
a_{5}  &  a_{4} & a_{3}
\end{array}
\right),
\ \ \cdots , \quad
T_{N} =
\det H.
\end{aligned}    
\end{equation}
Hence, the Routh-Hurwitz criterion can be summarized as $T_{0} = a_{0} > 0$, $T_{1} > 0$, $T_{2} > 0$, $\cdots$, $T_{N} > 0$. This is a rather convenient approach to assess the stability status of a given system, especially when working with high order polynomials, since it is not necessary to actually solve an otherwise convoluted expression that often does not admit analytical solutions. 

An equivalent set of constraints is given by the Liénard-Chipart criterion \cite{lienard1914signe}. Then, the roots of a polynomial given by Eq.~\eqref{eq:RH-polyn} are Hurwitz stable if, and only if, all coefficients and either all odd- or even-dimensional principal minors of the Hurwitz matrix are positive. Therefore, stability is guaranteed as long as 
\begin{equation}
\label{eq:LC-odd}
\begin{aligned}
&
a_{0} > 0, \quad a_{1} > 0, \quad \cdots, \quad a_{N} > 0, \text{ \textit{and} } T_{1} > 0, \quad T_{3} > 0, \quad T_{5} > 0, \quad \cdots, \quad T_{2 \lceil (N + 1)/2 \rceil - 1} > 0,
\end{aligned}    
\end{equation}
or
\begin{equation}
\label{eq:LC-even}
\begin{aligned}
&  a_{0} > 0, \quad a_{1} > 0, \quad \cdots, \quad a_{N} > 0, \text{ \textit{and} } T_{2} > 0, \quad T_{4} > 0, \quad T_{6} > 0, \quad \cdots, \quad T_{2\lceil N/2 \rceil} > 0,  
\end{aligned}    
\end{equation}
where $\lceil j \rceil$ denotes the largest integer not exceeding $j$. The Liénard-Chipart criterion demands calculating roughly half of the determinants required by the Routh-Hurwitz criterion. In this section, we shall carefully obtain a set of constraints for the transport coefficients that stem from the Liénard-Chipart criterion and are required to ensure that the stability of BDNK theory is \textit{always} fulfilled.

For the transverse modes, the dispersion relation, Eq.~\eqref{eq:BDNK-long-trans}, becomes
\begin{equation}
\bar{\lambda} \Upsilon^2 + \frac{4}{3} \Upsilon + \bar{\eta} k^2 = 0.
\end{equation}
In the context of the Liénard-Chipart criterion, stability dictates that $\bar{\lambda} > 0$ and $\bar\eta > 0$. The shear viscosity coefficient, $\eta$, is positive-definite, leading to a single non-trivial constraint, $\lambda > 0$, which is in agreement with the stability condition given by Eq.~\eqref{eq:BDNK-stab-long-small-k} and implies that $z>1$. 

Furthermore, the dispersion relation associated with the longitudinal modes becomes
\begin{eqnarray}
\label{eq:BDNK-long-disp-RH}
\Upsilon ^5 \bar\lambda (\bar\chi - \bar\xi) + \frac{1}{3} \Upsilon ^4 (3 \bar\lambda + 4 \bar\xi - 4 \bar\chi ) 
+
\frac{2}{3} \Upsilon ^3 \left[(2 \bar\eta + \bar\lambda ) (\bar\xi - \bar\chi) k^2 + 2 \right]
+
\frac{2}{9} k^2 \Upsilon ^2 (6 \bar\eta - 3 \bar\lambda + 2 \bar\xi + 6 \bar\varkappa - 2 \bar\chi )
\notag 
\\
+
\frac{1}{9} \Upsilon  \left[ \left(\bar\lambda (\bar\xi - \bar\chi) - 12 \bar\eta (\bar\lambda - \bar\varkappa)\right) k^4 + 4 k^2\right] 
+
\frac{1}{9} \left(4 \bar\varkappa - 3 \bar\lambda \right) k^4  = 0.   
\end{eqnarray}
The stability of the longitudinal modes is then guaranteed as long as the following inequalities are simultaneously satisfied:
\begin{subequations}
\label{eq:bdnk-stab-rh-transp}
\begin{align}
\bar\xi - \bar\chi &> 0, \\
12 \bar\eta (\bar\varkappa - \bar\lambda) + \bar\lambda (\bar\xi - \bar\chi) &> 0, \\
3 \bar\lambda^2 + 4 \eta  (\bar\xi - \bar\chi ) + \bar\lambda (\bar\xi  - \bar\chi - 3 \bar\varkappa) &> 0, \\
4 (\bar\xi - \bar\chi )^3 + 6 (2 \bar\eta +5 \bar\lambda - 4 \bar\varphi ) (\bar\xi - \bar\chi )^2 + 9 (\bar\lambda - \bar\varphi ) (-4 \bar\eta + 7 \bar\lambda -4 \bar\varphi ) (\bar\xi - \bar\chi ) + 27 \bar\lambda  (\bar\lambda - \bar\varphi ) [(\bar\lambda - \bar\varphi )- \bar\eta ] &> 0, \\
4 \bar\lambda (\bar\xi - \bar\chi)^4 + 36 \bar\lambda (\bar\lambda - \bar\varkappa) (\bar\xi - \bar\chi)^3 + 36 [\bar\lambda^3 - 6 \bar\varkappa \bar\lambda^2 + 3 \bar\varkappa^2 \bar\lambda - 3 (\bar\lambda - \bar\varkappa) \bar\eta \bar\lambda - 4 \bar\eta^2 (\bar\lambda - \bar\varkappa)] (\bar\xi - \bar\chi)^2  &\notag \\
+ 108 [\bar\lambda^4 - 3 \bar\varkappa \bar\lambda^3 + 3 \bar\varkappa^2 \bar\lambda^2 - \bar\varkappa^3 \bar\lambda + 3 \bar\eta \bar\varkappa \bar\lambda (\bar\lambda - \bar\varkappa)] (\bar\xi - \bar\chi) - 81 \bar\eta \bar\lambda^2 (\bar\lambda - \bar\varkappa) [\bar\lambda - \bar\varkappa + 4 (\bar\xi - \bar\chi)] &>0,
\end{align}
\end{subequations}
which stem from Eq.~\eqref{eq:LC-even}. In terms of the matching parameters $q$, $s$ and $z$, the stability conditions become
\begin{subequations}
\label{eq:bdnk-stab-param-lc}
\begin{align}
q + s - 3 &> 0, \label{eq:bdnk-stab-param-rh-first} \\
q + s - \frac{7z - 19}{z-1} &> 0, \\
q + s - \frac{z (8-z) + 5}{z + 3} &> 0, \\
(q+s)^3 + (4 z-15) (q+s)^2 + \left(4 z^2-51 z+107\right) (q+s)+ z^3-25 z^2+161 z-245 &> 0, \\
(q+s)^4 + 3 (z-8) (z-1) (q+s)^3 + \left[3 z^2 - 51 z - \frac{16(z-4)}{z-1} - 12 (z-4) + 210 \right] (q+s)^2 + \notag \\
\left[ z^3 - 30 z^2 - 12 (z - 4) z + 273 z + \frac{96(z-4)}{z-1} + 120 (z - 4) - 784 \right] (q+s)  +  \notag \\
- 3 z^3 - 4(z - 1) z^2 + 63 z^2 - \frac{144(z - 4)}{z-1} -268 (z - 4)
+ 1029 + 56 (z - 4) z - 441 z &> 0. \label{eq:bdnk-stab-param-rh-last}
\end{align}
\end{subequations}

\begin{figure}[ht]
\centering
\begin{subfigure}{0.5\textwidth}
  \includegraphics[width=\linewidth]{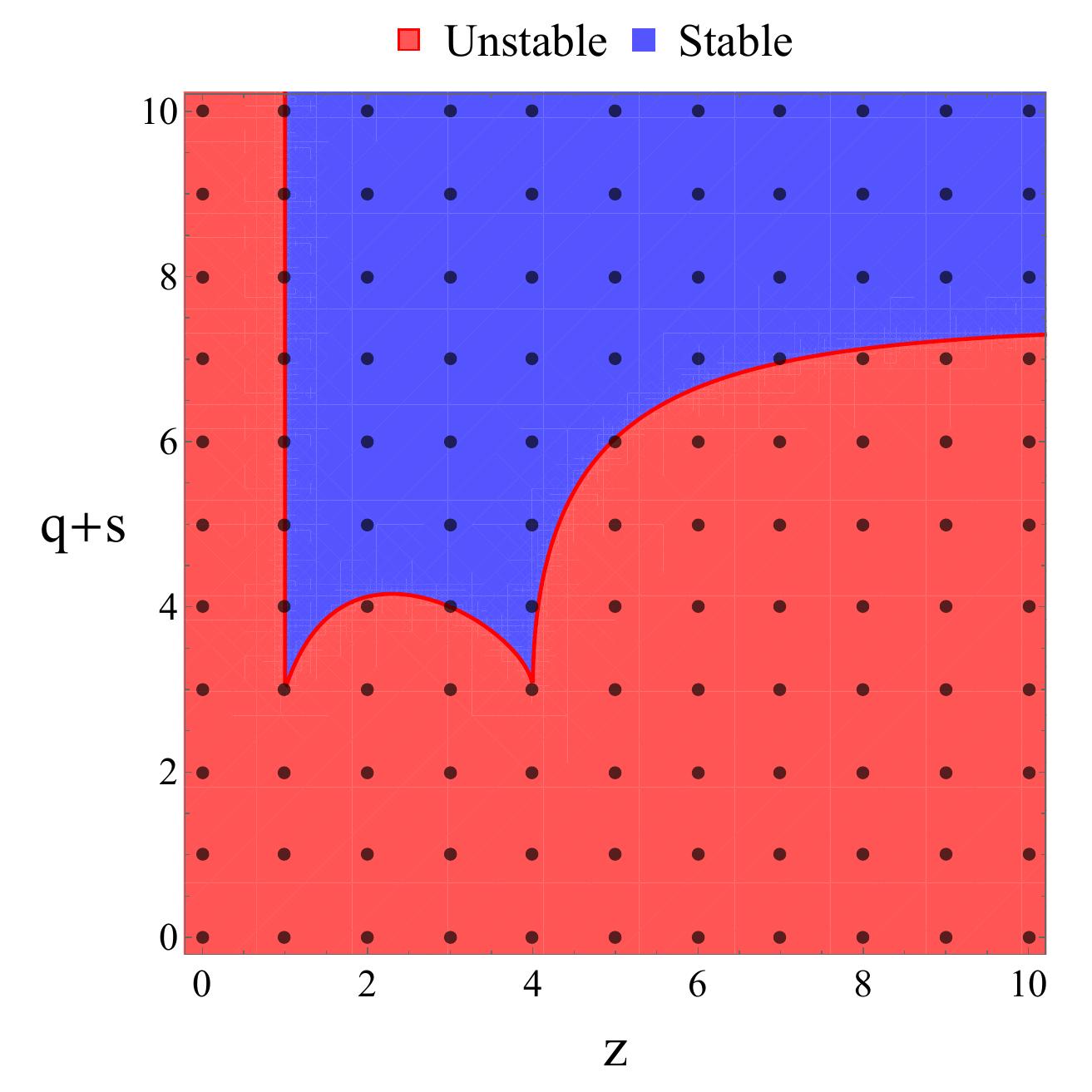}
  \caption{Stability and instability regions in matching parameter space according to the Liénard-Chipart criterion.}
  \label{fig:BDNK-stab-regions}
\end{subfigure}
\begin{subfigure}{0.5\textwidth}
  \includegraphics[width=\linewidth]{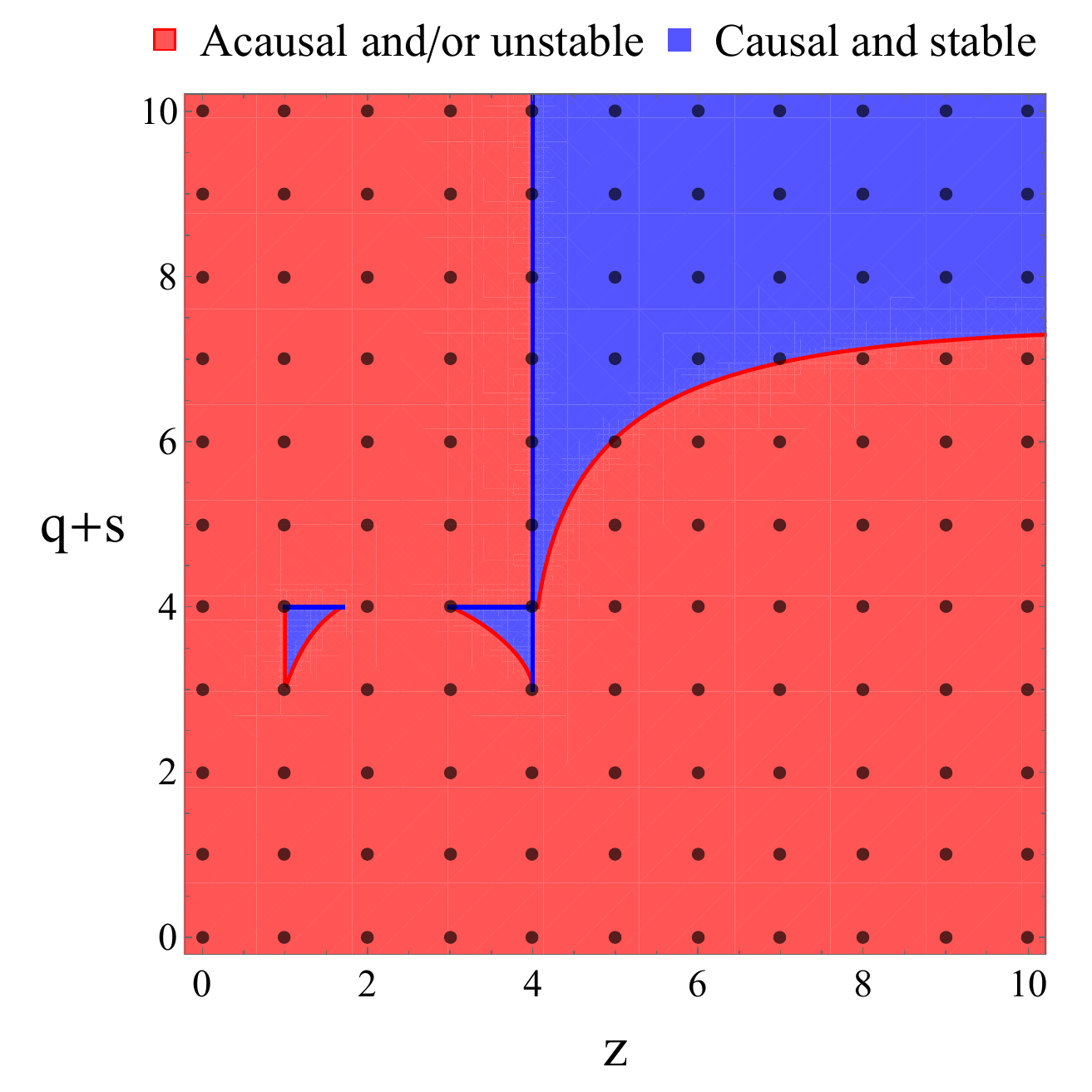}
 \caption{Causal \textit{and} stable regions in matching parameter space.}
  \label{fig:BDNK-caus-stab}
\end{subfigure}
\caption{(Color online) Left panel: regions in matching parameter space where all stability conditions stemming from the Liénard-Chipart are simultaneously satisfied (blue) or at least one of them is violated (red). Right panel: regions in parameter space where all stability and causality conditions are simultaneously satisfied (blue) or at least one of them is violated (red).}
\label{fig:BDNK-caus-and-stab}
\end{figure}

In Fig.~\ref{fig:BDNK-stab-regions}, we display the regions in parameter space in which BDNK theory is stable. This plot illustrates the values that the matching parameters $q$, $s$ and $z$ can assume in order for stability to be satisfied, i.e., where Eqs.~\eqref{eq:bdnk-stab-param-lc} are all satisfied. We remark that this approach provides a more restrict range for the values that $q$, $s$ and $z$ can assume in order to ensure the stability of theory, in comparison to the analysis developed in the last section. In Fig.~\ref{fig:BDNK-caus-stab}, we display the regions in parameter space where causality \textit{and} stability conditions, given by Eqs.~\eqref{eq:bdnk-causal-params} and \eqref{eq:bdnk-stab-param-lc}, respectively, are either all satisfied or at least one of them is violated. It can be readily seen that requiring causality conditions to be also fulfilled constrain even further the range of values allowed for the matching parameters. 

\begin{figure}[ht]
\begin{center}
\includegraphics[scale=0.35]{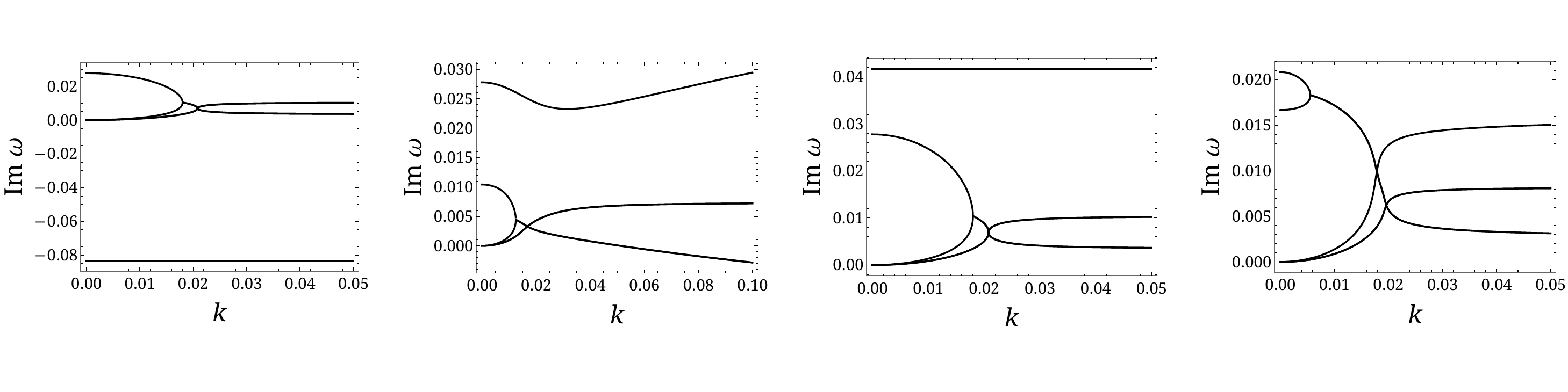} \\
\includegraphics[scale=0.35]{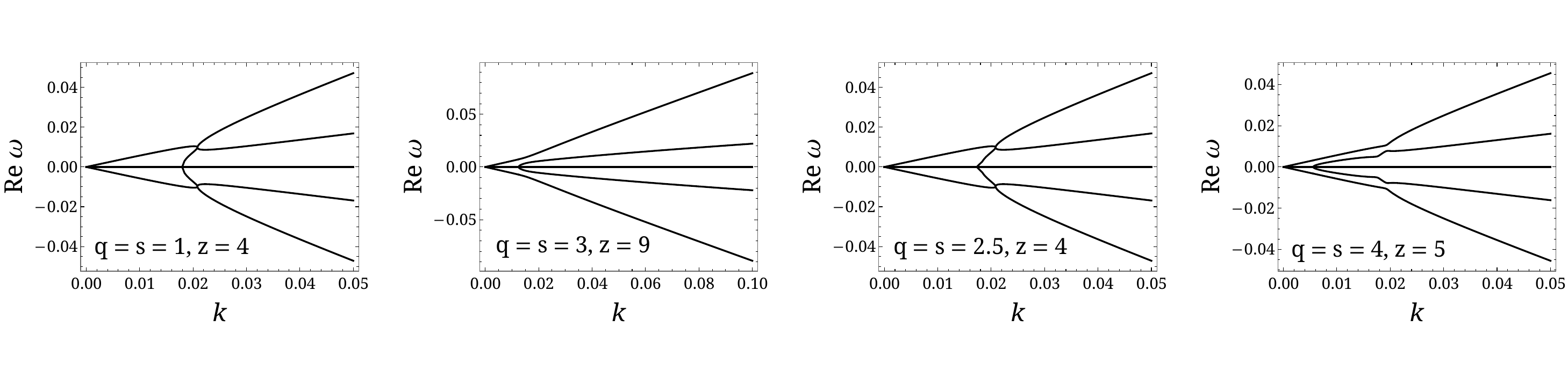}
\caption{Imaginary and real parts of the longitudinal modes of BDNK theory considering several values for the matching parameters $q$, $s$ and $z$.}
\label{fig:BDNK-long-modes}
\end{center}
\end{figure}

We display the imaginary and real parts of the longitudinal modes of BDNK theory considering several scenarios in Fig.~\ref{fig:BDNK-long-modes}. In the first column, we consider $q=s=1$ and $z=4$, violating stability in both approaches, see Fig.~\ref{fig:BDNK-stab-regions}. In the second column, we assume $q=s=3$ and $z=9$, hence obtaining modes that are asymptotically stable, while unstable according to the Liénard-Chipart criterion. In this case, we observe that, in fact, such criterion poses stronger -- in particular, \textit{necessary} -- constraints on stability and thus the modes are actually unstable. In the third column we assume $q=s=2.5$ and $z=4$, corresponding to a stable configuration according to both approaches. Last, in the fourth column, we consider $q=s=4$ and $z=5$, obtaining modes that are both causal and stable, see Fig.~\ref{fig:BDNK-caus-stab}.

\subsection{Transient fluid dynamics}

For the past decades, the linear stability and causality of transient theories of relativistic fluid dynamics have been extensively studied \cite{Hiscock:1983zz, Olson:1990rzl, Denicol:2008ha, Pu:2009fj, Brito:2020nou, Brito:2021iqr, Sammet:2023bfo}. The transport coefficients cannot assume arbitrary values if causality and stability are imposed. In particular, constraints on the relaxation times are crucial to ensure a subluminal signal propagation and are also essential for the linear stability of these theories. In Ref.~\cite{Rocha:2023hts}, the equations of motion for the diffusion current and shear-stress tensor for a system of classical massless particles self-interacting via a quartic potential were explicitly obtained from the Boltzmann equation within Landau matching conditions, Eqs.~\eqref{eq:IS-eoms}, and expressions for the transport coefficients were explicitly obtained in \textit{exact} form, Eqs.~\eqref{eq:IS-coeffs}. In this section, we analyze the causality and stability of this formulation.

In the linear regime, Eqs.~\eqref{eq:IS-eoms} reduce to
\begin{eqnarray}
\tau_\nu D_0 \Delta \nu^\mu + \Delta \nu^\mu &=& \varkappa_{n} \nabla^\mu_0 \Delta \alpha + \ell_{\nu\pi} \nabla_\lambda \Delta \pi^{\mu\lambda} + \mathcal{O}(\Delta^2), \\
\tau_\pi D_0 \Delta \pi^{\mu\nu} + \Delta \pi^{\mu\nu} &=& 2 \eta \Delta^{\mu\nu\alpha\beta}_0 \nabla_\alpha \Delta u_\beta + \ell_{\pi \nu} \Delta^{\mu\nu\alpha\beta}_0 \nabla_\alpha \Delta \nu_\beta + \mathcal{O}(\Delta^2).
\end{eqnarray}
Analogously to the previous analyses, we express these equations in terms of dimensionless variables in Fourier space
\begin{eqnarray}
\left( i\tau_\nu \Omega + 1 \right) \Delta \hat{\tilde{\nu}}^\mu &=& i \bar{\varkappa}_{n} \kappa^\mu \Delta \tilde{\alpha} + i \mathcal{L}_{\nu\pi} \kappa_\lambda \Delta \hat{\tilde{\pi}}^{\mu\lambda}, \label{eq:fourier-resc-IS-diff} \\
\left( i\tau_\pi \Omega + 1 \right) \Delta \hat{\tilde{\pi}}^{\mu\nu} &=& i 2 \bar{\eta} \left[ \frac{1}{2} \left( \kappa^\mu \Delta \tilde{u}^\nu + \kappa^\nu \Delta \tilde{u}^\mu \right) - \frac{1}{3} \Delta^{\mu\nu}_0 \kappa^\alpha \Delta \tilde{u}_\alpha \right] + i \mathcal{L}_{\pi\nu} \left[ \frac{1}{2} \left( \kappa^\mu \Delta \hat{\tilde{\nu}}^\nu + \kappa^\nu \Delta \hat{\tilde{\nu}}^\mu \right) - \frac{1}{3} \Delta^{\mu\nu}_0 \kappa^\alpha \Delta \hat{\tilde{\nu}}_\alpha \right],\label{eq:fourier-resc-IS-shear}
\end{eqnarray}
where $\mathcal{L}_{\nu\pi} = 3 \ell_{\nu\pi}/\beta$ and $\mathcal{L}_{\pi\nu} = \beta \ell_{\pi\nu}/3$, while the remaining rescaled coefficients are defined according to Eqs.~\eqref{eq:NS-resc-coeff} and \eqref{eq:BDNK-resc-coeff}.

\begin{figure}[ht]
\begin{center}
\includegraphics[scale=0.47]{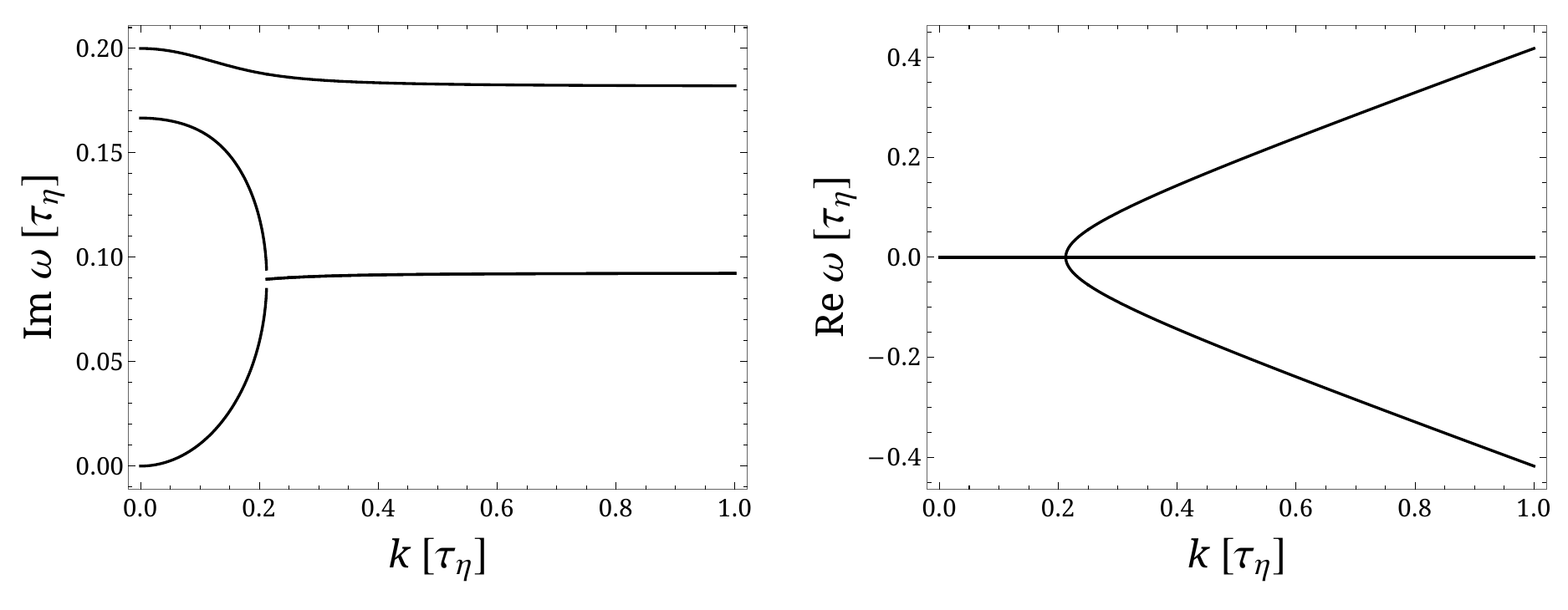}
\caption{Imaginary and real parts of the transverse modes of the transient theory for perturbations on a static background.}
\label{fig:IS-trans}
\end{center}
\end{figure}

The transverse projections of these equations are obtained by contracting Eq.~\eqref{eq:fourier-resc-IS-diff} with $\Delta_{\nu \alpha}^\kappa/\kappa$ and Eq.~\eqref{eq:fourier-resc-IS-shear} with $- \kappa_\mu \Delta_{\nu \alpha}^\kappa/\kappa$, respectively, leading to
\begin{eqnarray}
\left( i\tau_\nu \Omega + 1 \right) \Delta \hat{\tilde{\nu}}^\alpha_\bot + i \mathcal{L}_{\nu\pi} \kappa \Delta \hat{\tilde{\pi}}^{\alpha}_\bot &=& 0, \\
\left( i\tau_\pi \Omega + 1 \right) \Delta \hat{\tilde{\pi}}^{\alpha}_\bot - i \bar{\eta} \kappa \Delta \tilde{u}^\alpha_\bot - \frac{i}{2} \mathcal{L}_{\pi\nu}\kappa \Delta \tilde{\nu}^\alpha_\bot &=& 0.
\end{eqnarray}
These equations are then plugged into Eq.~\eqref{eq:linear-hydro-eqs-fourier-trans} and can be cast as
\begin{equation}
\left(
\begin{array}{ccc}
 \frac{4}{3} \Omega & -\kappa  & 0 \\
 0 & i \kappa  \mathcal{L}_{\nu \pi} & i \Omega  \tau _{\nu }+1 \\
 -i \bar\eta  \kappa  & i \Omega  \tau _{\pi}+1 & -\frac{1}{2} i \kappa  \mathcal{L}_{\pi \nu} \\
\end{array}
\right)
\left(
\begin{array}{c}
\Delta \tilde{u}^\mu_\bot \\
\Delta \hat{\tilde \pi}^\mu_\bot \\
\Delta \hat{\tilde \nu}^\mu_\bot
\end{array}
\right)
= 0,
\end{equation}
leading to the following dispersion relation
\begin{equation}
\frac{4}{3} \tau_{\nu} \tau_{\pi} \Omega^3
- \frac{4i}{3} \left(\tau_\nu + \tau_{\pi}\right) \Omega ^2
+ \frac{1}{3} \left[ \left( 2 \mathcal{L}_{\nu\pi} \mathcal{L}_{\pi\nu} -3 \bar\eta \tau_\nu \right) \kappa ^2 - 4\right] \Omega + i \bar\eta  \kappa ^2 = 0.
\end{equation}
Once again, we assume perturbations on a fluid at rest. In this scenario, in the small wave number limit, the transverse modes read
\begin{equation}
\omega = \frac{i}{\tau_\nu} + \mathcal{O}(k^2), \ \
\omega = \frac{i}{\tau_\pi} + \mathcal{O}(k^2), \ \ 
\omega = \frac{3 i \bar\eta }{4} k^2 + \mathcal{O}(k^3).
\end{equation}
Therefore, stability implies that $\tau_\nu > 0$, $\tau_\pi > 0$ and $\eta > 0$. These transport coefficients are, in fact, positive-definite, see Eqs.~\eqref{eq:IS-coeffs}, thus rendering the transverse modes stable in this regime. Moreover, in the large wave number limit, these modes behave as
\begin{equation}
\omega = \pm \frac{k}{2} \sqrt{\frac{3 \bar\eta \tau_\nu - 2 \mathcal{L}_{\nu \pi} \mathcal{L}_{\pi \nu}}{\tau_\nu \tau_\pi}} + \mathcal{O}(k^0), \ \ 
\omega = \frac{3 i \bar\eta}{3 \bar\eta \tau_\nu -2 \mathcal{L}_{\nu \pi} \mathcal{L}_{\pi \nu}} + \mathcal{O}\left( \frac{1}{k} \right).
\end{equation}
Stability implies that the first mode must be purely real, otherwise the negative solution would be unstable and therefore the term inside the square root has to be positive. In fact, since the relaxation times and shear viscosity are positive-definite and the product between the coupling terms is negative, see Eqs.~\eqref{eq:IS-coeffs}, stability is always guaranteed for all transverse modes in the large wave number limit. In addition, causality implies that
\begin{equation}
\lim_{k\rightarrow \infty }\left\vert\frac{\partial\mathrm{Re} \, \, \omega}{\partial k}\right\vert \leq1 \ \
\Longrightarrow \ \
3 \bar\eta \tau_\nu - 4 \tau_\nu \tau_\pi - 2 \mathcal{L}_{\nu \pi } \mathcal{L}_{\pi\nu} \leq 0, \label{eq:IS-trans-caus}
\end{equation}
which is indeed satisfied by the transport coefficients given in Eqs.~\eqref{eq:IS-coeffs}. Therefore, we conclude that the transverse modes are linearly causal and stable. In Fig.~\ref{fig:IS-trans}, we display the real and imaginary parts of the transverse modes of the transient theory with the frequency and wave number being expressed in units of $\tau_\eta \equiv 3 \bar\eta/4$. 

The final step is to look at the longitudinal modes. In particular, in order to obtain the longitudinal projections of Eqs.~\eqref{eq:fourier-resc-IS-diff} and \eqref{eq:fourier-resc-IS-shear}, we contract them with $- \kappa_\mu/\kappa$ and $\kappa^\mu \kappa_\nu/\kappa^2$, respectively,
\begin{subequations}
\label{eq:IS-long-dissip}
\begin{align}
\left( i\tau_\nu \Omega + 1 \right) \Delta \hat{\tilde{\nu}}_\| &= i \bar{\varkappa}_{n} \kappa \Delta \tilde{\alpha} - i \mathcal{L}_{\nu\pi} \kappa \Delta \hat{\tilde{\pi}}_\|, \\
\left( i\tau_\pi \Omega + 1 \right) \Delta \hat{\tilde{\pi}}_\| &= \frac{4}{3}i \bar\eta \kappa \Delta\tilde{u}_\| + \frac{2}{3} i \mathcal{L}_{\pi\nu} \kappa \Delta \hat{\tilde{\nu}}_\|.
\end{align}
\end{subequations}
Using Eqs.~\eqref{eq:e-and-n-as-alpha-beta}, we express the longitudinal projections of the conservation laws, Eqs.~\eqref{eq:linear-hydro-eqs-fourier-long}, as
\begin{subequations}
\label{eq:IS-long-cons}
\begin{align}
\Omega \left( \Delta\tilde\alpha - \frac{3}{\beta_0} \Delta\beta \right) - \kappa \Delta \tilde{u}_{\|} - \kappa \Delta\hat{\tilde{\nu}}_{\|} &= 0, \\
\Omega \left( \Delta\tilde\alpha - \frac{4}{\beta_0} \Delta\beta \right) - \frac{4}{3} \kappa \Delta \tilde{u}_{\|} &= 0, \\
\frac{4}{3} \Omega \Delta \tilde{u}_{\|} - \frac{1}{3} \kappa \left( \Delta\tilde\alpha - \frac{4}{\beta_0} \Delta\beta \right) -  \kappa \Delta\hat{\tilde{\pi}}_{\|} &= 0.
\end{align}
\end{subequations}
Then, Eqs.~\eqref{eq:IS-long-dissip} and \eqref{eq:IS-long-cons} can be expressed as
\begin{equation}
\left( 
\begin{array}{ccccc}
\Omega & - \frac{3}{\beta_0} \Omega & - \kappa & - \kappa & 0  \\ 
\Omega & - \frac{4}{\beta_0} \Omega & - \frac{4}{3} \kappa & 0 & 0  \\ 
-\frac{1}{3} \kappa & \frac{4}{3\beta_0} \kappa & \frac{4}{3} \Omega & 0 & - \kappa \\
- i \bar\varkappa_{n} \kappa & 0 & 0 & i \tau_\nu \Omega + 1 & i \mathcal{L}_{\nu\pi} \kappa \\
0 & 0 & - \frac{4}{3} i \bar\eta \kappa & - \frac{2}{3} i \mathcal{L}_{\pi \nu} \kappa & i \tau_\pi \Omega + 1
\end{array}%
\right) \left( 
\begin{array}{c}
\Delta \tilde\alpha \\ 
\Delta \tilde\beta \\
\Delta \tilde{u}_{\|} \\
\Delta \hat{\tilde{\nu}}_\| \\
\Delta \hat{\tilde{\pi}}_\|
\end{array}%
\right)=0,
\end{equation}
leading to the following dispersion relation
\begin{eqnarray}
&& 9 \tau _{\nu } \tau _{\pi } \Omega ^5
- 9i \left(\tau_{\nu } + \tau_{\pi}\right) \Omega ^4
- 3 \left[ \left(3 \bar\eta   \tau _{\nu } +  \tau _{\nu} \tau _{\pi} + 12 \bar\varkappa_{n}  \tau _{\pi} - 2  \mathcal{L}_{\nu \pi} \mathcal{L}_{\pi \nu }\right)\kappa ^2+3\right] \Omega ^3 \notag \\
&+& 3i \left(3 \bar\eta + \tau _{\nu } + \tau _{\pi} + 12 \bar\varkappa_{n} \right) \kappa^2 \Omega ^2 
+ \left[ \left(36 \bar\eta   \bar\varkappa_{n} + 12  \bar\varkappa_{n}  \tau _{\pi} - 2  \mathcal{L}_{\nu \pi } \mathcal{L}_{\pi \nu }\right) \kappa^2 + 3 \right] \kappa^2  \Omega  
-12 i \kappa ^4 \bar\varkappa_{n}
= 0.
\end{eqnarray}
In the small wave number, the longitudinal modes behave as
\begin{equation}
\omega =  \frac{i}{\tau_\nu} + \mathcal{O}(k^2), \ \ 
\omega = \frac{i}{\tau_\pi} + \mathcal{O}(k^2), \ \ 
\omega = \pm \frac{1}{\sqrt{3}} k + \mathcal{O}(k^2), \ \ 
\omega = 4 i \bar\varkappa_{n} k^2 + \mathcal{O}(k^3),
\end{equation}
and are identically stable, see Eqs.~\eqref{eq:IS-coeffs}. In the large wave number limit, on the other hand, the longitudinal modes become
\begin{subequations}
\begin{align}
\omega &= \pm k \sqrt{\frac{\tau _{\nu } \left(3 \bar\eta +\tau _{\pi}\right)+12 \bar\varkappa_{n}  \tau _{\pi } -2 \mathcal{L}_{\nu \pi } \mathcal{L}_{\pi \nu } \pm \sqrt{\left[\tau _{\nu } \left(3 \bar\eta +\tau _{\pi}\right)+12 \bar\varkappa_{n}  \tau _{\pi }-2 \mathcal{L}_{\nu \pi } \mathcal{L}_{\pi \nu }\right]^2 - 8 \tau _{\nu } \tau _{\pi } \left[6 \bar\varkappa_{n}  \left(3 \bar\eta +\tau _{\pi }\right)-\mathcal{L}_{\nu \pi } \mathcal{L}_{\pi \nu }\right]}}{6 \tau _{\nu } \tau _{\pi }}}, \label{eq:IS-sound-mode} \\
\omega &= \frac{6 i \bar\varkappa_{n} }{6 \bar\varkappa_{n}  \left(3 \bar\eta +\tau _{\pi }\right)-\mathcal{L}_{\nu \pi } \mathcal{L}_{\pi \nu }} + \mathcal{O}\left( \frac{1}{k} \right).
\end{align}    
\end{subequations}
First, stability implies that the term inside the inner square root must be real, otherwise there would be a mode with a negative imaginary part, hence imposing the condition
\begin{equation}
\label{eq:IS-long-stab-1}
\left[\tau_\nu \left(3 \bar\eta + \tau_\pi\right) + 12 \bar\varkappa_{n}  \tau_\pi - 2 \mathcal{L}_{\nu \pi} \mathcal{L}_{\pi \nu}\right]^2 - 8 \tau_\nu \tau_\pi \left[6 \bar\varkappa_{n}  \left(3 \bar\eta +\tau_\pi \right) - \mathcal{L}_{\nu \pi } \mathcal{L}_{\pi \nu}\right]\geq 0,
\end{equation}
which is indeed satisfied by the transport coefficients given in Eq.~\eqref{eq:IS-coeffs}. Likewise, the term inside the outer square root must also be real. Since the relaxation times are positive-definite, this implies that the numerator must be positive as well, which is guaranteed by
\begin{equation}
\label{eq:IS-long-stab-2}
6 \bar\varkappa_{n} \left(3 \bar\eta + \tau_\pi \right) - \mathcal{L}_{\nu \pi} \mathcal{L}_{\pi \nu} \geq 0. 
\end{equation}
As a matter of fact, the product of the coupling terms is negative, while $\varkappa_{n}$, $\eta$ and $\tau_\pi$ are positive-definite, see Eqs.~\eqref{eq:IS-coeffs}, hence the inequality above is trivially satisfied. This also guarantees the stability of the remaining mode.

In addition to having exponentially decreasing amplitudes, these modes must also propagate subluminally, i.e., they must be stable as well as causal. Causality is then guaranteed by the following condition
\begin{equation}
\label{eq:IS-long-caus-1}
3 \left(3 \bar\eta - 2 \tau_\pi \right) \left(4 \bar\varkappa_{n} -\tau_{\nu }\right)+4 \mathcal{L}_{\nu \pi} \mathcal{L}_{\pi \nu } \geq 0.
\end{equation}
In particular, in deriving this condition, the transport coefficients still must satisfy an additional constraint, given by
\begin{equation}
\label{eq:IS-long-caus-2}
5 \tau_\nu \tau_\pi +2 \mathcal{L}_{\nu\pi} \mathcal{L}_{\pi\nu} - 3 \bar\eta  \tau_\nu - 12 \bar\varkappa_{n} \tau _{\pi } \geq 0.
\end{equation}
Finally, using the transport coefficients summarized in Eqs.~\eqref{eq:IS-coeffs}, we verify that the stability and causality conditions given by Eqs.~\eqref{eq:IS-long-stab-1}-\eqref{eq:IS-long-caus-2} are simultaneously satisfied. Therefore, we conclude that transient fluid dynamics for a system of weakly self-interacting classical massless particles is linearly causal and stable. In particular, the propagating modes given in Eq.~\eqref{eq:IS-sound-mode} reduce to
\begin{equation}
\omega = \pm 0.456 \, k, \ \ 
\omega = \pm 0.755 \, k.
\end{equation}

\begin{figure}[ht]
\begin{center}
\includegraphics[scale=0.47]{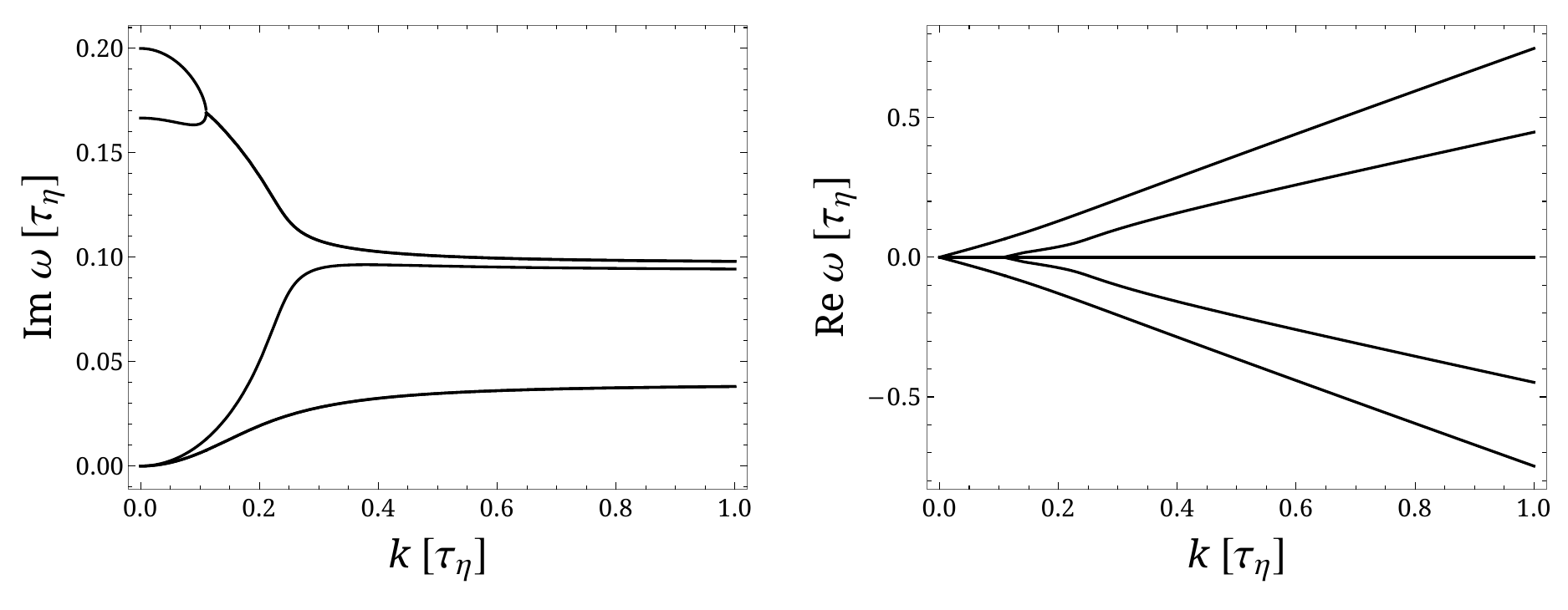}
\caption{Imaginary and real parts of the longitudinal modes of the transient theory for perturbations on a static background.}
\label{fig:IS-long}
\end{center}
\end{figure}

For the sake of illustration, in Fig.~\ref{fig:IS-long}, we display the real and imaginary parts of the longitudinal modes of the transient theory. Once again, it is readily seen that all modes are causal as well as stable.

\section{Conclusions}
\label{sec:conc}

We have analyzed the causality and stability of several fluid-dynamical formulations for a system of weakly self-interacting classical massless particles derived in Ref.~\cite{Rocha:2023hts}. The Navier-Stokes theory was shown to be linearly acausal and unstable for arbitrary matching conditions, as both transverse and longitudinal modes have a diffusion-like behavior in the large wave number limit. In this regime, the dispersion relation is parabolic, leading to an additional (unstable) mode when an observer other than one sitting on the rest frame of the system is taken into account. In particular, it is not possible to tune the transport coefficients in order for causality and stability to be fulfilled, even by considering different matching conditions. We remark that this result was expected, since the acausal and unstable character of Navier-Stokes theory is an already well-known issue for Eckart and Landau matching conditions \cite{Hiscock:1985zz}.

The Bemfica-Disconzi-Noronha-Kovtun theory is a formalism that describes the dissipative currents as constitutive relations that includes not only space-like derivatives, but also time-like ones. We have shown that this leads to non-parabolic dispersion relations, and therefore such formulation is not affected by the aforementioned problems of Navier-Stokes theory. As a matter of fact, causality and stability in this case depend on the value of the transport coefficients, which in turn depend on the choice of matching conditions. Therefore, causality and stability of the BDNK theory are strongly matching-dependent. In this work, we have obtained the complete set of \textit{necessary} conditions for the transport coefficients (and, consequently, for the matching parameters) that must be satisfied in order for causality and stability to be simultaneously fulfilled. In particular, we remark that these constraints prohibit the so-called exotic Eckart matching \cite{Bemfica:2019knx,Rocha:2022ind}, for which $\nu^{\mu} \equiv 0$ ($z=0$, in our notation), which lies in the unstable region in Figs.~\ref{fig:BDNK-caus-and-stab}.

The transient fluid-dynamical formulation derived in Ref.~\cite{Rocha:2023hts} was restricted to Landau matching conditions. We have explicitly calculated the asymptotic behavior of the modes and concluded that, if the transport coefficients derived in exact form in Ref.~\cite{Rocha:2023hts} are employed, causality and stability are always fulfilled, and all modes are well-behaved for any value of wave number. This result is in agreement with previous analyses of Israel-Stewart theory \cite{Pu:2009fj, Brito:2020nou, Sammet:2023bfo}.

\section*{Acknowledgments}

C. V. P. B. is partly funded by Coordenação de Aperfeiçoamento de Pessoal de Nível Superior (CAPES), Finance Code 001, Award No. 88881.722616/2022-01 and by Conselho Nacional de Desenvolvimento Científico e Tecnológico (CNPq), Grant No. 140453/2021-0.  G. S. R. was funded by Vanderbilt University, by CAPES, Finance code 001, Award No. 88881.650299/2021-01 and also by CNPq, Grant No.~142548/2019-7.  G. S. D. also acknowledges CNPq as well as Fundação Carlos Chagas Filho de Amparo à Pesquisa do Estado do Rio de Janeiro (FAPERJ), Grant No.~E-26/202.747/2018.

\appendix

\section{Causality and stability analysis for Hilbert theory}
\label{apn:hilb}

In Appendix B of Ref.~\cite{Rocha:2023hts}, the microscopic derivation of transport coefficients for the Hilbert theory has been performed. Historically, the Hilbert expansion \cite{hilbert1912begrundung,grad1958principles} was the first systematic procedure for deriving hydrodynamic equations of motion from kinetic theory. However, since it did not lead to Navier-Stokes theory, such an approach is usually not discussed, even in non-relativistic contexts. Based on Hilbert's work, Enskog \cite{enskog1917kinetische} and Chapman \cite{chapman1916vi} proposed their expansion, which is indeed a resummation of Hilbert expansion. The relativistic counterpart of Hilbert expansion in generic matching conditions was discussed in Ref.~\cite{Rocha:2022ind}. In this Appendix, we discuss the causality status of such theory.

The main feature of the Hilbert equations of motion is that local conservation laws are obeyed independently at each order in the gradient expansion. In this case, the Euler equations are identically satisfied,  
\begin{subequations}
 \label{eq:euler-eqns-hilb}
\begin{align}
 \label{eq:hydro-EoM-n0-euler}
D n_{0} + n_{0} \theta &= 0, \\
\label{eq:hydro-EoM-eps-euler}
D\varepsilon_{0} + (\varepsilon_{0}+ P_{0}) \theta &= 0, \\
\label{eq:hydro-EoM-umu-euler}
(\varepsilon_{0} + P_{0})Du^{\mu} - \nabla^{\mu}P_{0}  &= 0,
\end{align}
\end{subequations}
and at first order, which is the order at which we will truncate, dissipative currents obey
\begin{subequations}
 \label{eq:hydro-EoMs-hilb}
\begin{align}
 \label{eq:hydro-EoM-n0-hilb}
D n_{(1)} + n_{(1)} \theta + \partial_{\mu} \nu^{\mu}_{(1)} &= 0, \\
\label{eq:hydro-EoM-eps-hilb}
D \varepsilon_{(1)} + (\varepsilon_{(1)} + \Pi_{(1)}) \theta - \pi^{\mu \nu}_{(1)} \sigma_{\mu \nu} + \partial_{\mu}h^{\mu}_{(1)} + u_{\mu} Dh^{\mu}_{(1)} &= 0, \\
\label{eq:hydro-EoM-umu-hilb}
(\varepsilon_{(1)} + \Pi_{(1)}) Du^{\mu} - \nabla^{\mu}\Pi_{(1)} + h^{\mu}_{(1)} \theta + h^{\alpha}_{(1)} \Delta^{\mu \nu} \partial_{\alpha}u_{\nu} +  \Delta^{\mu \nu} Dh_{(1)\nu} + \Delta^{\mu \nu} \partial_{\alpha}\pi^{\alpha}_{(1)\nu} &= 0,
\end{align}
\end{subequations}
where the thermodynamic fields $\alpha$, $\beta$, and $u^{\mu}$ enter as non-constant coefficients, since they are solutions of Eqs.~\eqref{eq:euler-eqns-hilb}. The subscript $(1)$ denotes that the dissipative currents are only considered at first order in gradients. The equations of motion above are supplemented by the following constraints/constitutive relations, valid for a gas composed of massless particles,
\begin{subequations}
\label{eq:constrains-H}
\begin{align}
&
\label{eq:constrains-H-1}
\Pi_{(1)} = \frac{1}{3} \varepsilon_{(1)},
\\
&
\label{eq:constrains-H-2}
\nu^{\mu}_{(1)} - \frac{\beta}{4} h^{\mu}_{(1)} 
=
\Tilde{\kappa}_{H}\nabla^{\mu} \alpha
,  
\\ 
&
\label{eq:constrains-H-3}
\pi^{\mu \nu}_{(1)} = 2 \eta \sigma^{\mu \nu}, 
\end{align}    
\end{subequations}
where the Hilbert transport coefficients, computed in Ref.~\cite{Rocha:2023hts}, read 
\begin{equation}
\begin{aligned}
&
\Tilde{\kappa}_{H}= \frac{3}{g \beta^{2}},
\\
&
\eta = \frac{48}{g\beta^{3}}.
\end{aligned}    
\end{equation}

Considering linear perturbations around global equilibrium, so that
\begin{equation}
\begin{aligned}
&
\varepsilon = \varepsilon_{0} + \varepsilon_{(1)} \rightarrow \varepsilon_0 + \Delta\varepsilon_0 + \Delta\delta\varepsilon, \hspace{.1cm}
n = n_{0} + n_{(1)} \rightarrow n_0 + \Delta n_0 + \Delta\delta n,
\hspace{.1cm} \Pi_{(1)} \rightarrow \Delta \Pi,
\\
&
\hspace{.1cm} u^\mu \rightarrow u^\mu_0 + \Delta u^\mu, \hspace{.1cm}
\nu^\mu_{(1)} \rightarrow \Delta \nu^\mu, \hspace{.1cm}
h^\mu_{(1)} \rightarrow \Delta h^\mu, \hspace{.1cm}
\pi^{\mu\nu}_{(1)} \rightarrow \Delta \pi^{\mu\nu},
\end{aligned}
\end{equation}
Eqs.~\eqref{eq:euler-eqns-hilb}-\eqref{eq:constrains-H} become
\begin{subequations}
 \label{eq:euler-eqns-pert}
\begin{align}
 \label{eq:hydro-EoM-n0-euler-pert}
D_{0} \Delta n + n_{0} \nabla_{0\mu}\Delta u^{\mu} &= 0 + \mathcal{O}(\Delta^{2}), \\
D_{0}\Delta \varepsilon + \frac{4}{3}\varepsilon_{0} \nabla_{0\mu}\Delta u^{\mu} &= 0 + \mathcal{O}(\Delta^{2}), \\
\label{eq:hydro-EoM-umu-euler-pert}
\frac{4}{3}\varepsilon_{0} D_{0}\Delta u^{\mu} - \nabla^{\mu}_{0} \Delta P &= 0 + \mathcal{O}(\Delta^{2}),
\\
 D_{0}\Delta\delta n + \nabla_{0 \mu}\Delta \nu^{\mu} &= 0 + \mathcal{O}(\Delta^{2}), \\
\label{eq:hydro-EoM-eps-hilb-apn}
D_{0} \Delta\delta \varepsilon + \nabla_{0 \mu}\Delta h^{\mu} &= 0 + \mathcal{O}(\Delta^{2}), \\
\label{eq:hydro-EoM-umu-hilb-apn}
- \frac{1}{3} \nabla^{\mu}\Delta\delta \varepsilon + \Delta^{\mu \nu}_{0} D_{0} \Delta h_{\nu}
+
\Delta^{\mu \nu}_{0} 2 \eta
\nabla_{\alpha}\Delta\sigma^{\alpha}_{\ \nu}
&= 0 + \mathcal{O}(\Delta^{2}), \\
\Delta \nu^{\mu} - \frac{\beta}{4} \Delta h^{\mu} 
& =
\Tilde{\kappa}_{H} \nabla_{0}^{\mu} \Delta \alpha,
\end{align}
\end{subequations}
where Eqs.~\eqref{eq:constrains-H-1} and \eqref{eq:constrains-H-3} have been explicitly substituted in Eqs.~\eqref{eq:hydro-EoM-eps-hilb} and \eqref{eq:hydro-EoM-umu-hilb}. In Fourier space, the equations of motion above read   
\begin{subequations}
\begin{align}
i \Omega \Delta \Tilde{n} + n_{0} i\kappa_{\mu}\Delta \Tilde{u}^{\mu} &= 0 + \mathcal{O}(\Delta^{2}), \\
\label{eq:hydro-EoM-eps-euler-pert}
i \Omega \Delta \Tilde{\varepsilon} + \frac{4}{3}\varepsilon_{0} i\kappa_{\mu}\Delta \Tilde{u}^{\mu} &= 0 + \mathcal{O}(\Delta^{2}), \\
\frac{4}{3}\varepsilon_{0} i \Omega \Delta \Tilde{u}^{\mu} - \frac{i}{3} \kappa^{\mu} \Delta \Tilde{\varepsilon} &= 0 + \mathcal{O}(\Delta^{2}),
\\
\label{eq:hydro-EoM-n0-hilb-apn}
 i \Omega \Delta\delta \Tilde{n} + i\kappa_{\mu}\Delta \Tilde{\nu}^{\mu} &= 0 + \mathcal{O}(\Delta^{2}), \\
i \Omega \Delta\delta \Tilde{\varepsilon} + i\kappa_{\mu} \Delta \Tilde{h}^{\mu} &= 0 + \mathcal{O}(\Delta^{2}), \\
- \frac{1}{3} i\kappa^{\mu}\Delta\delta \Tilde{\varepsilon} +  i \Omega \Delta \Tilde{h}^{\mu}
-
\eta \left[ - \kappa^{2} \Delta \Tilde{u}^{\mu}  + \frac{1}{3}  (\kappa_{\alpha} \Delta \Tilde{u}^{\alpha}) \kappa^{\mu} \right]
&= 0 + \mathcal{O}(\Delta^{2}), \\
\Delta \Tilde{\nu}^{\mu} - \frac{\beta}{4} \Delta \Tilde{h}^{\mu} 
& =
\frac{3}{g \beta^{2}} i \kappa^{\mu} \Delta \Tilde{\alpha},
\end{align}
\end{subequations}
which imply in the following expression for the transverse modes,
\begin{equation}
\label{eq:M-Uid-Udiss}
\begin{aligned}
&\left(
\begin{array}{ccc}
i \Omega & 0 & 0 \\
\Bar{\eta} \kappa^{2} & 0 & i \Omega \\
0 & 1 & - \frac{3}{4} 
\end{array}
\right)
\left(
\begin{array}{c}
\Delta \Tilde{u}_{\perp}^{\mu} \\
\Delta \Hat{\Tilde{\nu}}_{\perp}^{\mu} \\
\Delta \Hat{\Tilde{h}}_{\perp}^{\mu}
\end{array}
\right)
=
0 .
\end{aligned}    
\end{equation}
Equating the determinant of the above-defined matrix to zero leads to $\Omega^{2} = 0$, which implies that these modes are non-propagating. On the other hand, the longitudinal modes are expressed as
\begin{equation}
\label{eq:long-modes-hilb}
\begin{aligned}
 &
 \left(
 \begin{array}{cc}
   \mathcal{M}^{\text{id}}_{3 \times 3}   &  \mathbf{0}_{3 \times 4} \\
   \mathcal{M}^{\text{id-diss}}_{4 \times 3}   &  \mathcal{M}^{\text{diss}}_{4 \times 4}
 \end{array}
 \right) 
 \left(
 \begin{array}{cc}
 \Delta U_{||}^{\text{id}}     \\
 \Delta U_{||}^{\text{diss}}      
 \end{array}
 \right)
 =
 0,
\end{aligned}
\end{equation}
where we define variable vectors $\Delta U_{||}^{\text{id}} \equiv \left( \Delta \Hat{\Tilde{n}}, \Delta \Hat{\Tilde{\varepsilon}}, \Delta \Tilde{u}_{||} \right)^{T}$, and $\Delta U_{||}^{\text{diss}} = \left(\Delta \delta \Hat{\Tilde{n}}, \Delta \delta \Hat{\Tilde{\varepsilon}}, \Delta \Hat{\Tilde{\nu}}_{||}, \Delta \Hat{\Tilde{h}}_{||}\right)^{T}$, and the matrices 
\begin{equation}
\begin{aligned}
&
 \mathcal{M}^{\text{id}}_{3 \times 3}
 =
 \left( 
\begin{array}{ccc}
i \Omega & 0 & - i \kappa
 \\
0 & i \Omega & - \frac{4}{3}i \kappa  \\
0 & - \frac{i}{3} \kappa  & \frac{4}{3} i \Omega
\end{array}
 \right),
\quad
 \mathcal{M}^{\text{id-diss}}_{4 \times 3} 
 =
  \left( 
\begin{array}{ccc}
0 & 0 & 0 \\
0 & 0 & 0 \\
0 & 0 &  -\frac{4}{3} \Bar{\eta} \kappa^{2} \\
i \kappa \Bar{\varkappa}_{H} &
\frac{4}{3}i \kappa \Bar{\varkappa}_{H} & 0 
\end{array}
 \right),
 \\
 &
 \mathcal{M}^{\text{diss}}_{4 \times 4}
 = 
\left( 
\begin{array}{cccc}
i \Omega & 0 & - i \kappa & 0\\
0 & i \Omega & 0 & - i \kappa\\
0 & - \frac{i \kappa}{3} & 0 & i \Omega \\
0 & 0 & 1 & - \frac{3}{4}
\end{array}
 \right).
\end{aligned}    
\end{equation}
Due to the structure of the Hilbert equations of motion, which contain the Euler equations, the longitudinal modes do not depend on the elements of the matrix $\mathcal{M}^{\text{id-diss}}_{4 \times 3}$. Indeed, the dispersion relation defined by Eq.~\eqref{eq:long-modes-hilb} leads to  
\begin{equation}
\begin{aligned}
&  
\det \left(
 \begin{array}{cc}
   \mathcal{M}^{\text{id}}_{3 \times 3}   &  \mathbf{0}_{3 \times 4} \\
   \mathcal{M}^{\text{id-diss}}_{4 \times 3}   &  \mathcal{M}^{\text{diss}}_{4 \times 4}
 \end{array}
 \right) 
=
(\det \mathcal{M}^{\text{id}}_{3 \times 3}) (\det \mathcal{M}^{\text{diss}}_{4 \times 4}) = -\frac{4}{3} \left[\Omega\left(\Omega^{2} -  \frac{1}{3} \kappa^{2}\right)\right]^{2} = 0, 
\end{aligned}   
\end{equation}
which leads to double-degenerate non-propagating modes ($\Omega = 0$) and two sound modes ($\Omega = (1/\sqrt{3}) \kappa$), which propagate with velocity $v_{\mathrm{sound}} = 1/\sqrt{3} \simeq 0.577$. One of such propagating modes stems from the ideal equations of motion, contained in the matrix $\mathcal{M}^{\text{id}}_{3 \times 3}$, and the other from the dissipative part, contained in the matrix $\mathcal{M}^{\text{diss}}_{4 \times 4}$. Since the speed of sound of both modes is smaller than that of light and larger than zero, Hilbert theory is causal and stable. For massive particles, this degeneracy of the velocity of the two propagating modes is broken, i.e.~the sound velocities become different \cite{Rocha:2023ths}, but both remain subluminal and positive, thus causality is preserved.  

\bibliographystyle{apsrev4-1}
\bibliography{refs}

\end{document}